\documentclass[a4paper,11pt]{article}
\pdfoutput=1 % if your are submitting a pdflatex (i.e. if you have
\usepackage{jinstpub} % for details on the use of the package, please
\usepackage{indentfirst}
\usepackage{subcaption}
\usepackage{lineno}
\usepackage{rotating}
\usepackage{colortbl}
\usepackage{arydshln}
\usepackage{float}
\newcommand{\e}[1]{\times 10^{#1}}

\title{\boldmath Electromagnetic Shower Reconstruction and Energy Validation with Michel Electrons and $\pi^0$ Samples for the Deep-Learning-Based Analyses in MicroBooNE}

\begin{document}

% Authors in alphabetical order
%\author[t]{P.~Abratenko}   % only for papers using MCS!
\author[gg]{P.~Abratenko}
\author[n]{R.~An}
\author[d]{J.~Anthony}
\author[r]{L.~Arellano}
\author[ff]{J.~Asaadi}
\author[dd]{A.~Ashkenazi}
\author[k]{S.~Balasubramanian}
\author[k]{B.~Baller}
\author[t]{C.~Barnes}
\author[w]{G.~Barr}
% ...

% Institutions in alphabetical order
% ...
\affiliation[d]{University of Cambridge, Cambridge CB3 0HE, United Kingdom}
% ...
\affiliation[k]{Fermi National Accelerator Laboratory (FNAL), Batavia, IL 60510, USA}
% ...
\affiliation[n]{Illinois Institute of Technology (IIT), Chicago, IL 60616, USA}
% ...
\affiliation[r]{The University of Manchester, Manchester M13 9PL, United Kingdom}
% ...
\affiliation[t]{University of Michigan, Ann Arbor, MI, 48109, USA}
% ...
\affiliation[w]{University of Oxford, Oxford OX1 3RH, United Kingdom}
% ...
\affiliation[dd]{Tel Aviv University, Tel Aviv, Israel, 69978}
% ...
\affiliation[ff]{University of Texas, Arlington, TX, 76019, USA}
\affiliation[gg]{Tufts University, Medford, MA, 02155, USA}
% ...

  \emailAdd{microboone\_info@fnal.gov}
\date{}

\abstract{
This article presents the reconstruction of the electromagnetic activity from electrons and photons (showers) used in the MicroBooNE deep learning-based low energy electron search. The reconstruction algorithm uses a combination of traditional and deep learning-based techniques to estimate shower energies. We validate these predictions using two $\nu_{\mu}$-sourced data samples: charged/neutral current interactions with final state neutral pions and charged current interactions in which the muon stops and decays within the detector producing a Michel electron.  Both the neutral pion sample and Michel electron sample demonstrate agreement between data and simulation. Further, the absolute shower energy scale is shown to be consistent with the relevant physical constant of each sample: the neutral pion mass peak and the Michel energy cutoff.
}

\keywords{Neutrino detectors; Noble liquid detectors (scintillation, ionization, double-phase); Time projection Chambers (TPC); Pattern recognition, cluster finding, calibration and fitting methods}

\arxivnumber{2110.11874} % only if you have one

\collaboration[c]{The MicroBooNE Collaboration}

\maketitle
\flushbottom

\section{Introduction}

The primary goal of the MicroBooNE experiment is investigate the anomalous excess of electron-like events observed in the MiniBooNE detector \cite{miniboone}. 
% The excess of events in MiniBooNE peaks at reconstructed neutrino energy below 500 MeV while the neutrino flux peaks at higher energy. Therefore, the anomaly is often called the Low Energy Excess (LEE).  
The anomaly is an excess of single electromagnetic showers with a peak shower energy in the 200-500 MeV observed in the MiniBooNE Cherenkov detector. The anomaly is here referred to as the Low Energy Excess (LEE). 
% That excess is interpreted as electron neutrino charged-current quasielastic (CCQE) scattering,  $\nu_e+n \rightarrow e^- +p $, where the proton is below Cherenkov threshold.
The MicroBooNE collaboration has developed several analyses designed to isolate LEE events, where event here refers to one detector readout record.  The reconstruction tools presented here make up the shower reconstruction phase of the deep learning (DL)-based analysis \cite{DLPRD}.  A preliminary version of the (DL)-based analysis utilized in this study can be seen in ref.\cite{Moon:2020kjr}. This analysis isolates events with 1 electron and 1 proton ($1e1p$) in the final state.  The neutrino energy is reconstructed as:
\begin{equation}
    E_\nu= K_{p} + K_{e} + M_{e} + M_{p} - (M_{n} - B),
\end{equation}
where $K$ indicates kinetic energy, $M$ is mass, $B$ is nuclear binding energy, and $e$ and $p$ indices indicate the electron and proton, respectively.  The proton kinetic energy is reconstructed from the length of the track and the known energy deposited per unit length in liquid argon \cite{3Dreco}.   This article describes the reconstruction of the electron  kinetic energy, which, for our $\nu_e$ signal of interest, ranges from 35 MeV to 1200 MeV.   Across this range, the topology of the deposited electron energy  changes from track-like at low energy to shower-like at high energy where photon radiation dominates. In this article, all electron energy deposits will be called showers despite the variety of topologies. The discussion will include both electrons and photons.

This article reports the method of electromagnetic shower reconstruction used in the DL-based analyses. The  $\nu_e$ simulation-derived shower-charge-to-energy conversion value is presented in section \ref{sec:showerreco}. We then use two low-energy samples to validate the shower reconstruction.  The first sample consists of photons produced by neutral pions ($\pi^0$) decays and is described in section \ref{sec:pions}.  The second sample consists of Michel electrons produced when a stopped muon from a $\nu_\mu$  charged current (CC) interaction decays and is described in section \ref{sec:michels}.  We provide data-simulation comparison plots for each sample that we use to validate the relative shower energy scale between data and simulation. The use of the $\nu_e$ simulation-derived  energy calculation is then further verified on data by measuring the agreement of data and simulation with two well-measured physical quantities: the $\pi^0$ invariant mass and the cutoff of the Michel energy spectrum.

\section{Preliminary Reconstruction of Events in the Deep Learning Analysis}
\label{sec:initialreco}

The MicroBooNE detector is a 2.6 m $\times$ 2.3 m $\times$ 10.4 m Liquid Argon (LAr) Time Projection Chamber (TPC) filled with 85 metric tons of LAr \cite{detectorpaper}.   The ionization electrons produced by charged particles in the event drift with a velocity of 0.1098~cm/$\mu$s through an applied potential of -70 kV to three wire planes \cite{SCEpub}.   The orientation of wires in the induction planes, $U$ and $V$, are $+60^\circ$ and $-60^\circ$ relative to vertical.   The collection plane, $Y$, has vertical wires.   The wire spacing in all planes is 0.3 cm.  The detector also contains a light collection system made up of 32 photomultiplier tubes that are used for triggering and initial event selection. The light collection system is not leveraged in the shower reconstruction work presented here.

The studies in this article use MicroBooNE data taken from 2016 to 2018 which were recorded over three run periods with $1.75\times 10^{20}$ protons-on-target (POT) in Run 1,  $2.70\times 10^{20}$ POT in Run 2 and $2.43\times 10^{20}$ POT in Run 3. There are various Monte Carlo (MC) simulation samples that are used to build the reconstruction algorithm. The simulated neutrino events are overlaid with off-beam cosmic muon data. These "overlay samples" are used throughout the rest of this article. They contain all types of simulated neutrino events expected in the given POT. There are three samples of this type, one corresponding to cosmic information collected in each of the three MicroBooNE  data runs used in the study which corresponds to $6.67\times10^{20}$ total POT. Additionally, for the $\pi^0$ study we incorporate a specialized overlay sample containing a larger number of events with a $\pi^0$ that decays to two photons in the final state (high POT $\pi^0$ sample). The incorporation of this sample allows for a reduction of the statistical uncertainty on the simulation. For the Michel study, we incorporate an overlay sample of neutrino events interacting outside of the MicroBooNE detector, as muons from these external $\nu_\mu$ CC interactions which enter the detector, come to a stop, and then decay provide an additional source of Michel electrons.

Various final state topologies occur at the neutrino energies observed in MicroBooNE. Those relevant to the DL-based LEE analysis have two particles attached to a vertex: $1e1p$ and $1\mu 1p$ events from $\nu_e$ and $\nu_\mu$ charged-current quasi-elastic or meson exchange current scattering, other types of $\nu_\mu$ induced events with only two particles reconstructed at the vertex (such as $\pi^0$ events with disconnected photons), and cosmic ray muons. The first steps of the ``low level reconstruction'' is to isolate the two-particle-vertices of interest.  The methods have been described elsewhere~\cite{Moon:2020kjr,3Dreco}; we briefly review them here.

\begin{figure}[H]
\centering
\begin{subfigure}[b]{.49\textwidth}
         \centering
         \includegraphics[width=\textwidth]{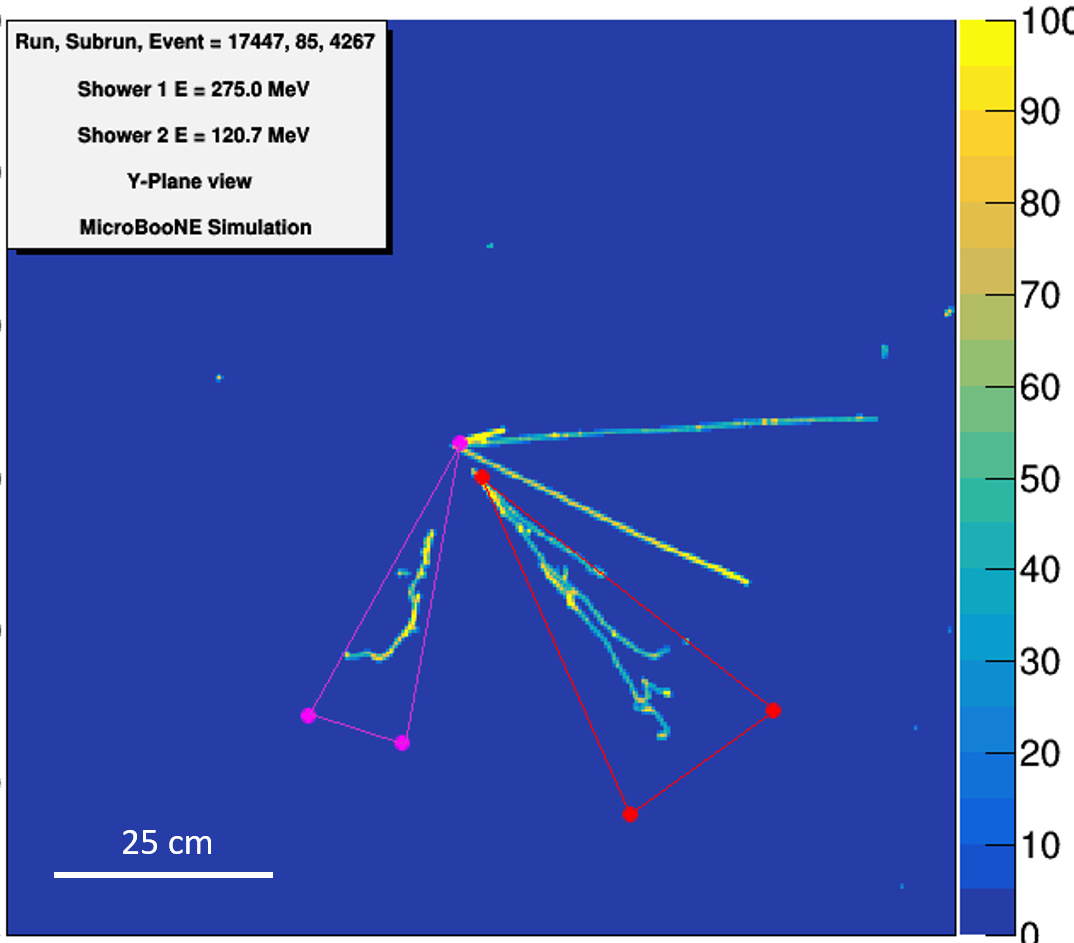}
         \caption{Q Image}
     \end{subfigure}
     
        \begin{subfigure}[b]{.49\textwidth}
         \centering
         \includegraphics[width=\textwidth]{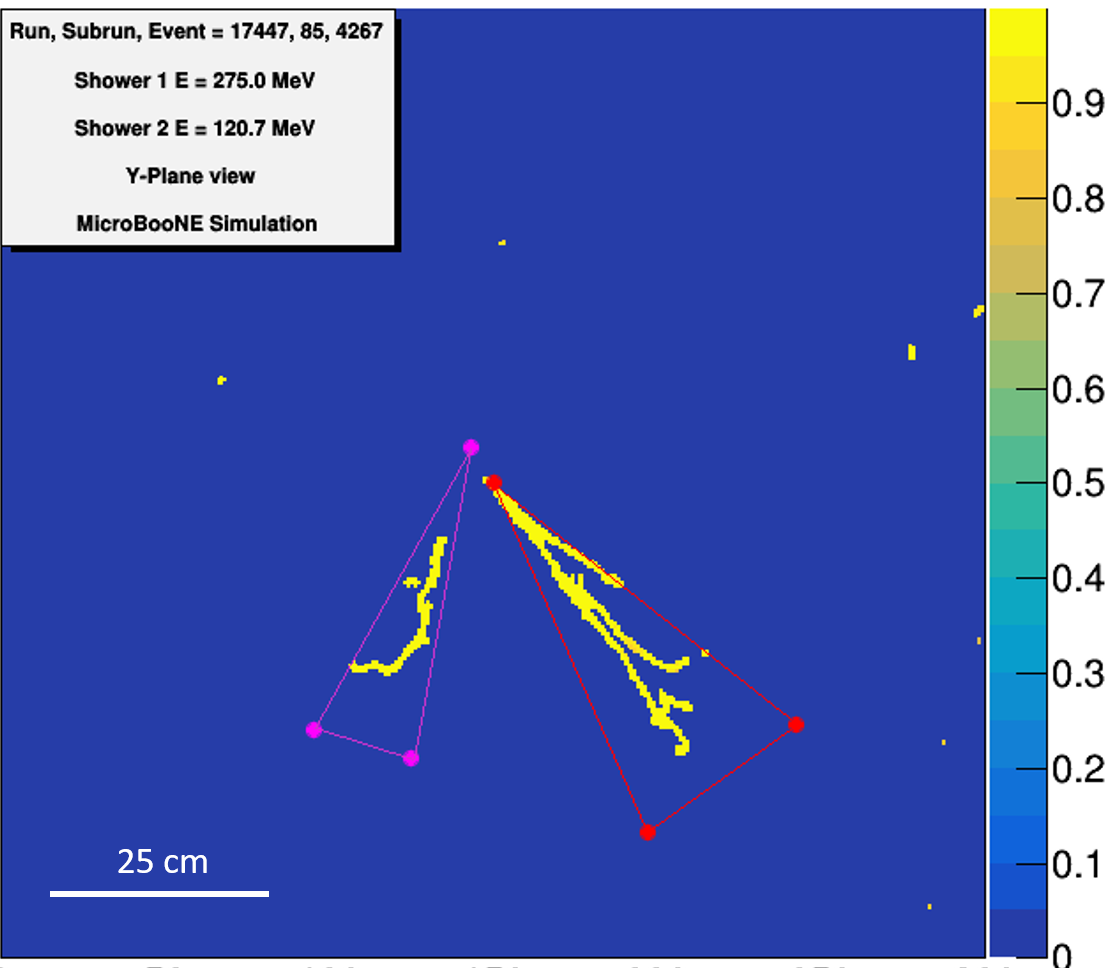}
         \caption{SparseSSNet Shower Image}
     \end{subfigure}

\caption{Event displays of a  simulated CC $\pi^0$ event. (a) shows the raw Q image and (b) shows the SparseSSNet shower score with track-like particles masked out. In both (a) and (b) the leading reconstructed photon is represented by the red triangle and the sub-leading reconstructed photon (shower 2) is represented by the magenta triangle.
}
\label{fig:eventdisp}
\centering
\end{figure}

Preparation for reconstruction begins with algorithms to tag and discard the charge associated with the cosmic rays \cite{WCTagger1,WCTagger2}. Next the waveform data from the wires in each plane are converted to ``images'' which are two-dimensional distributions with wire number along the $x$ axis and drift time along the $y$ axis.  The intensity of each image ``pixel'' is given by the integrated reconstructed charge waveform from six 0.5~$\mu$s TPC time increments after applying noise-filtering ~\cite{noisefilter} and signal processing ~\cite{signal1,deteff2}. The six 0.5~$\mu$s TPC time increments is comparable to the 3~mm wire spacing after accounting for the electron drift velocity.  The intensity of each pixel is referred to as "Q (charge)" in this article. An example event display of this data format is shown in figure \ref{fig:eventdisp} (a). The event display shown is of the collection plane ($Y$ plane) of a simulated CC $\pi^0$ event which passes the shower reconstruction stage. The z-axis represents the charge ($Q$) at each (wire,time) pixel. To make the deposited charge more clear in the display, $Q$ has been given a maximum value of 100.  In the event shown, the image is cropped centering at the neutrino interaction vertex. Shower reconstruction is also shown which will be discussed in section \ref{sec:showerreco}.

Any charge tagged as associated with a cosmic ray is removed at this stage. The $U$, $V$, and $Y$ images are then  passed into the deep learning convolutional neural net, called ``SparseSSnet,''which is a semantic segmentation algorithm that labels pixels as track-like or shower-like \cite{SparseSSnet}. Figure \ref{fig:eventdisp} (b) shows the SparseSSNet shower scores of the same event as (a) with the track-like particles masked out.

The next step is to identify a ``two particle'' vertex, followed by 3D reconstruction of each particle.   This process is described in detail for the $1 \mu 1p$ sample in ref.~\cite{3Dreco}, and the $1e1p$ reconstruction follows similar steps.
In short, the vertex algorithm searches for a characteristic ``vee'' shape where two particles meet at a vertex identifying cases where the particles are longer than 3 cm and the opening angle in at least one plane is greater than 10$^\circ$. Based on the SparseSSnet pixel tagging, the vee may be formed of a ``track-track'' pair or a ``shower-track'' pair~\cite{SparseSSnet}.
For any given vertex, more than one vee may be found if there are more than two particles emitted from the interaction point.   Also, more than one vertex may be found at different points in the same event.   An important example of this, used in this article, is the case of $1\mu 1p$ final state in which the muon decays to a Michel electron.   This results in one vertex at the interaction point and another at the decay point.   Because vertices are also found on cosmic rays, particularly those that stop and decay, many vertices are identified in any given event. All are passed to 3D reconstruction. At this stage in the reconstruction, multiple vertices can be reconstructed in each event. In later selections, a single vertex from the event will be chosen.

3D reconstruction of track-like objects (primarily protons and muons) is described in detail in ref.~\cite{3Dreco} and summarized here. The algorithm begins at the reconstructed vertex and follows ionization trails outward in 3D, clustering the charge into ``prongs.''   A prong is defined here as a collection of continuous charge, but it need not be a single line of charge. The prong may comprise connected branches of charge as will be the case for electromagnetic showers.    Each prong is assumed to come from one particle. In the case where two prongs are identified, we identify the proton as the prong having the higher average pixel-based ionization density.

A fiducial volume containment requirement is enforced on all prongs. This requirement uses the distance of a prong from the edge of the detector as the minimal distance from all the prong's 3D points to a detector edge. It is required that either the distance of both prongs is $>5$\,cm from the edge or else that the combined distance of both prongs is $>15$\,cm from the edge.

In order to associate a prong with the SparseSSnet identified pixels, the 3D prong is projected onto the 2D images described above.  The pixels in this projection are then matched to the prong. The kinetic energies of track-like prongs are calculated using the track lengths. The direction of track-like particles is also reconstructed as described in ref.~\cite{3Dreco}. Energy deposits from electrons and photons will suffer gaps due to radiated photons which the 3D track reconstruction cannot handle. Shower-like prongs therefore need a seperate reconstruction as described in section \ref{sec:showerreco}.

\section{Electromagnetic Shower Energy Reconstruction}
\label{sec:showerreco}

The  electromagnetic shower reconstruction algorithm is run to find any associated shower particles and reconstruct their kinetic energies once a candidate vertex is isolated.  The method described here builds on a previous MicroBooNE shower reconstruction described in ref.~\cite{ubshowerreco}. The algorithm described retains many vital features of the previous version, especially the use of a semantic segmentation neural network for pixel labeling. Important updates have been added including many simplifications of the algorithm made possible by the improved SparseSSnet.

The first step of the reconstruction is to mask the image to only use the pixels identified as shower by SparseSSNet. SparseSSNet outputs a value for each pixel indicating how likely it is that the pixel is part of a shower. This value, called shower score, falls between 0.0 and 1.0 where 1.0 indicates a shower-like pixel. Pixels with a shower score of $>$0.5 and intensity $>$10 $Q$ (charge) counts are kept, all other pixels are masked out for the rest of the shower reconstruction.  A threshold of 10 $Q$ is used to remove wire noise. This cut is standard among all tools used in this analysis. The value was chosen based on the distribution of $Q$ from minimum ionizing particles (MIPs),which peaks at $\approx40$\ $Q$.  A template isosceles triangle is then placed with its apex at the reconstructed vertex position, pointing in the positive wire direction. The triangle is optimized to choose the shower direction, length, and opening angle for which the triangle contains the most pixels with non-zero charge. These parameters each start at the minimum value shown in Table~\ref{showerreco_params}. In order to allow for showers that are detached from the vertex, a gap parameter is introduced allowing the triangle to start further from the vertex.  

Once a first shower candidate has been found by the reconstruction algorithm, the pixels found in the shower are  masked out. If the total amount of charge remaining passing the cuts of a shower score of $>$0.5 and intensity $>10$ $Q$(charge)  is $>5000$ $Q$, the second shower algorithm is run on the masked image. The total range of allowed values for each of the template triangle parameters is shown in Table~\ref{showerreco_params}. In this table "first shower" refers to the shower found in the first pass of the shower reconstruction which is aimed at finding showers near the reconstructed vertex. "Second shower" refers to the shower found on the second run of the algorithm and has expanded parameters to search for detached showers as described in the following paragraphs. The parameters in each case are optimized sequentially in the order of direction, gap size, opening angle, and length.

\begin{table}[htb]
    \centering {
       \caption{ Range of parameters for the shower reconstruction algorithm. Parameters are changed in the second shower search to allow for the capture of showers detached from the vertex.   
    \label{showerreco_params}}
    \begin{tabular}{|l|c|c|}
    \hline
    & Minimum value & Maximum value \\
    \hline
    Direction        & 0 degrees & 360 degrees\\
    Opening angle         & 17 degrees & 75 degrees \\
    Length (first shower)     & 3 cm & 35 cm \\
    Length (second shower)      & 3 cm & 60 cm \\
    Gap Size (first shower) & 0 cm & 17 cm \\
    Gap Size (second shower) & 0 cm & 90 cm \\
    \hline
    \end{tabular}}

\end{table}

Figure \ref{fig:eventdisp} shows an example display demonstrating the 2D shower reconstruction on a simulated CC $\pi^0$ event. The SparseSSNet shower-like particles are shown in (b) with the track-like particles masked out. The final optimized showers are shown in red (first shower) and magenta (second shower). In this example, the algorithm found the proper gap of the first shower, but not the second shower. This is acceptable as gap size is not a value that is utilized in any other part of the DL analysis. 
The reconstructed energy of each shower is also reported. Energy reconstruction of electromagnetic showers is a crucial component of the LEE analysis this work supports, which is designed to measure electrons from CC neutrino interactions over a a broad energy range from 35 to 1200 MeV. This analysis aims to develop and validate a energy reconstruction procedure for showers in this broad energy range. To determine the energy of each shower, the charge of all shower pixels enclosed in the $Y$-view triangle is integrated. This total shower charge will hereby be denoted by $Q_{sh}$. We use the $Y$-view, which is the collection plane,  because it has the highest signal-to-noise ratio of the three planes \cite{noisefilter}.

\begin{figure}[h]
\includegraphics[width=\textwidth]{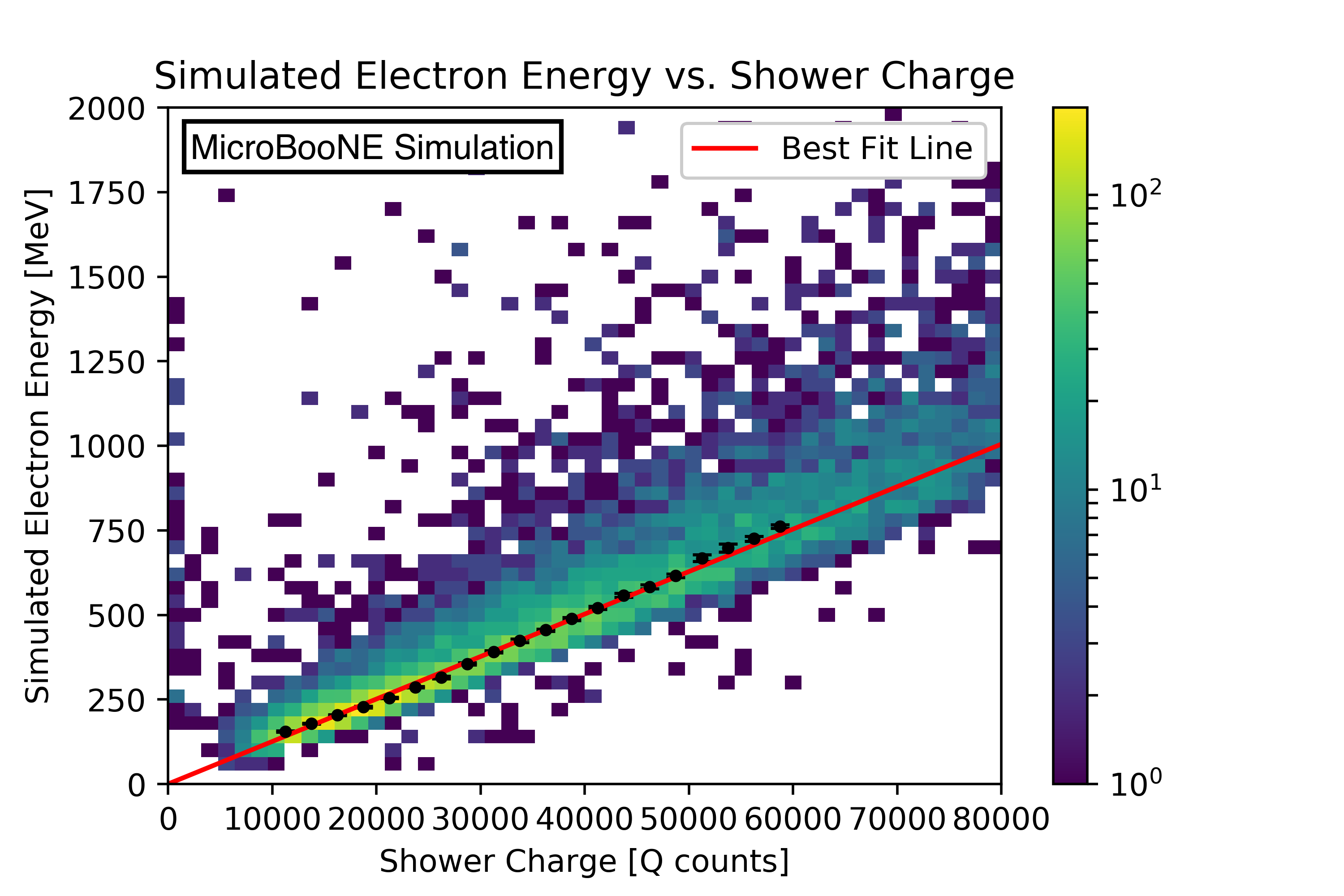}
\caption{Simulated electron energy vs $Q_{sh}$ for a sample of generated $1e1p$ events. The linear fit is used in the shower energy calculation.
}
\label{ADCtoMeV}
\centering
\end{figure}

\begin{figure}[h]
\includegraphics[width=\textwidth]{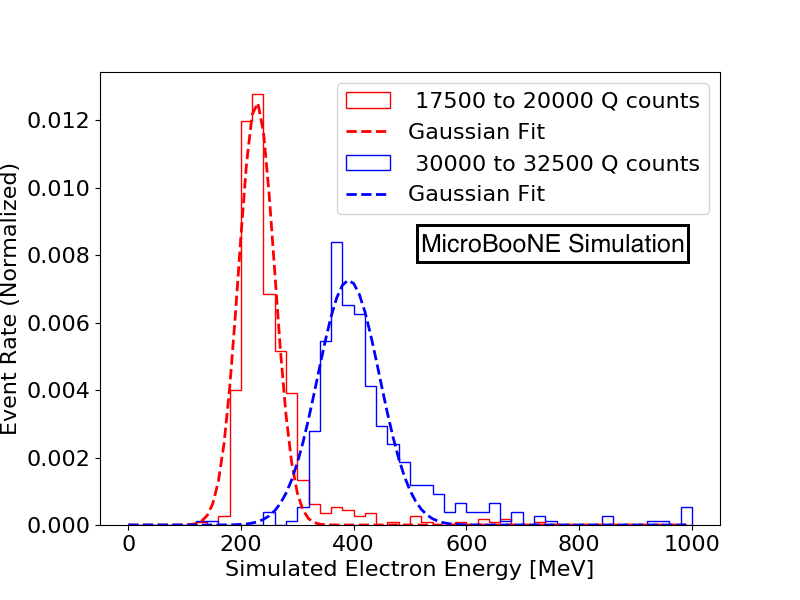}
\caption{Example distributions of simulated electron energies (solid lines) and corresponding Gaussian distributions (dashed lines) within two different shower charge sum ranges. The peaks of the Gaussian fits are used to generate the black points in figure ~\ref{ADCtoMeV}.
}
\label{fig:gaus_example}
\centering
\end{figure}

To convert the reconstructed charge to energy, the $Q_{sh}$ in a sample of simulated events is compared to the generated energy of the electrons in the events. A $Q_{sh}$-to-MeV conversion line is determined as described below and shown in figure~\ref{ADCtoMeV}. As the focus of the larger analysis is $1e1p$ events, a sample of simulated $1e1p$ events are used. The simulated electron energy is plotted versus the reconstructed $Y$-view total shower charge sum for events in a $\nu_e$ simulation sample selected by the MicroBooNE DL-based $1e1p$ analysis. In each vertical bin, the peak is found with the help of a Gaussian distribution (represented by the black points in figure~\ref{ADCtoMeV}). The edge bins which have smaller statistics are excluded. Two examples of the Gaussian fits are shown in figure ~\ref{fig:gaus_example}.While the Gaussian fits capture the bulk of simulated events well, one can see a tail in each distribution out to higher simulated electron energies which is not captured by the Gaussian. This is expected, as these tails correspond to electron showers which are not fully reconstructed (i.e., showers which pass through unresponsive wires or exit the active volume). The points are then fit to a line. The slope of this line is used for the $Q_{sh}$-to-MeV conversion and is referred to in the rest of this article as $m_{e^-}$.  While a Gaussian is not a perfect fit to the data in, the use of $m_{e^-}$, derived from these points, will be validated in section \ref{sec:pions} and section \ref{sec:michels}. The resulting equation is:

\begin{equation}
\label{eq:orig_calib}
     ~\text{Electron: }E~\text{[MeV]} = (1.26 \pm 0.01 \e{-2})  \times Q_{sh}~\text{[$Q$ Counts]}.
\end{equation}
where the error corresponds to the uncertainty on the linear fit, which represents the statistical error on the simulated electron sample used for the fit.

Various detector effects could cause this fit value to change.
%The different detector systematic variations that will be investigated  are discussed in more detail Section \ref{sec:pi0ADCtoMeVCheck}. 
Specifically the amount of detected energy will depend on the position, amount of energy deposited, and the orientation of the particle's trajectory with respect to the wires~\cite{signal1,deteff2,wirecalib}. Additionally, the value of $m_{e^-}$ derived here assumes good argon purity and could be affected by periods of low purity. Potential systematic uncertainty could be introduced by these effects. The results shown in Sec \ref{sec:finalcompare} give an estimate of the size of the detector effect.

Using this shower energy calculation, we look at the energy resolution for a sample of  simulated CC $\nu_e$ events containing an electron and no final state $\pi^0$. This isolates electrons from photons which will be discussed further in section \ref{pi0id}. The following selection criteria are used:

\begin{enumerate}
    \item Reconstructed vertex is less than 5 cm from simulated neutrino interaction vertex;
    \item One simulated electron contained in event;
    \item No final state $\pi^0$;
    \item One reconstructed shower; and
    \item 1e1p Boosted Decision Tree (BDT) score is greater than 0.7 \cite{Moon:2020kjr}.
\end{enumerate}

\begin{figure}[h]
\includegraphics[width=\textwidth]{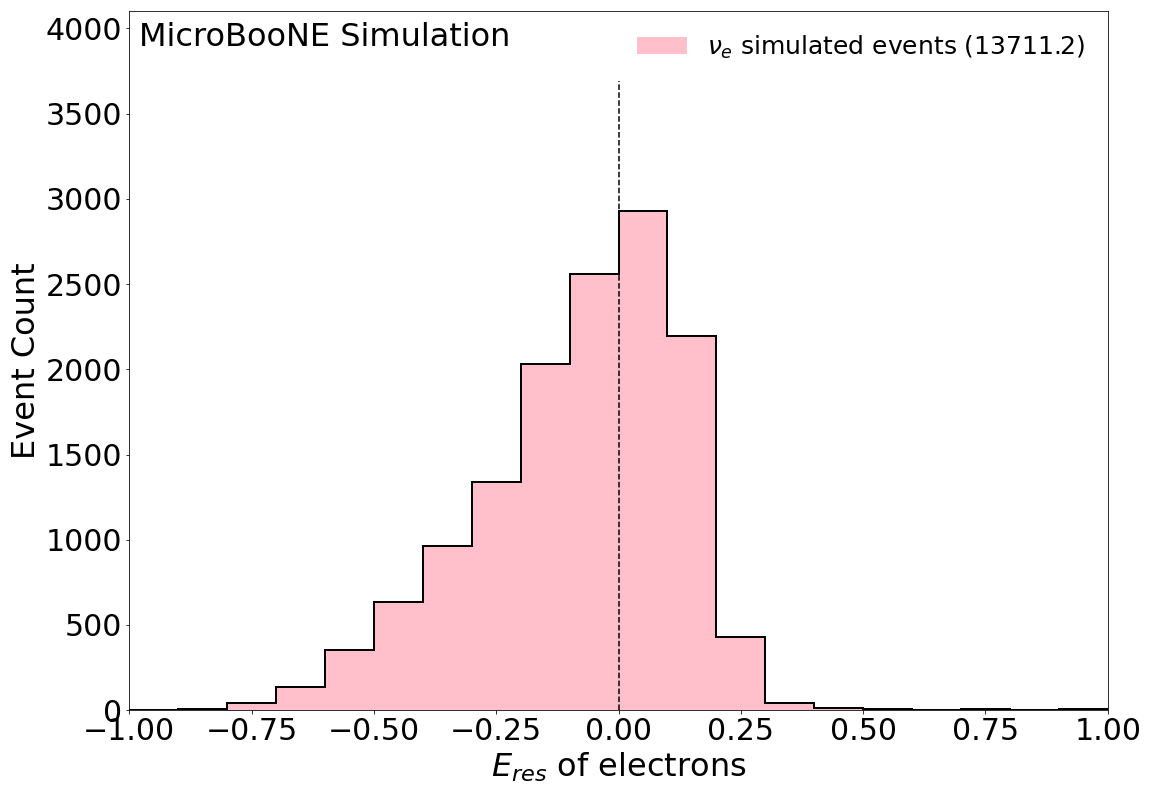}
\caption{The energy resolution for a sample of simulated electrons as described by eq. \ref{eq:res}. The $y$ axis has the raw number of simulated events without scaling. The dashed vertical line is included at $E_{res}=0.0$ for reference.}
\label{fig:electronres}
\centering
\end{figure}

The energy resolution, defined here as:
\begin{equation}
\label{eq:res}
    E_{res} = \frac{E_{reco} - E_{sim}}{E_{sim}}
\end{equation}
for this sample of simulated events is seen in figure \ref{fig:electronres}. The mean is at -0.07 and the RMS is 0.22. 
%A Gaussian fit to the distribution gives a bias of $4\%$ and a resolution of $19\%$.

The shower energy reconstruction presented here will be utilized and validated in section \ref{sec:pions} and section \ref{sec:michels}. As stated earlier, the shower energy is calculated using a MicroBooNE simulation sample comprising of simulated neutrino events overlay with off-beam cosmic ray data. It is therefore notable that the energy calculated with this sample works well when applied to data samples as shown in the next sections. It will be seen that the $\pi^0$ and Michel $e^-$ samples both have good data/simulation agreement using eq. (\ref{eq:orig_calib}). A further test is done for each sample that finds the charge-to-energy conversion factor that gives the best agreement to known physical values (the $\pi^0$ rest mass and the Michel electron spectrum cutoff). This test shows great data/simulation agreement with each sample. The results are also comparable to $m_{e^-}$ given in  eq. (\ref{eq:orig_calib}), even at different energy scales, validating the use of this linear conversion factor.

\section{Neutral Pion Sample}
\label{sec:pions}

The first sample used to analyze the performance of the shower reconstruction is a sample of $\pi^0$ events. The identification and reconstruction of $\pi^0$ events is presented. This is followed by both a verification of the $Q_{sh}$-to-MeV conversion value ($m_{e^-}$) agreement to data and simulation, and a verification of agreement between data and simulation which is accomplished by using a well-measured physical quantity: the $\pi^0$ invariant mass (135 MeV). 

\subsection{Identification and Reconstruction}
\label{pi0id}

The reconstruction of $\pi^0$ events relies on the shower reconstruction described in section \ref{sec:showerreco} and introduces  3D shower reconstruction. The starting point is either a track-track or track-shower vertex.  The shower reconstruction is then applied as discussed above,  centering on clusters of electromagnetic charge, but not requiring that the charge be attached to the starting vertex.  This leads to the bulk of $\pi^0$ events selected for this analysis matching two specific topologies.  The most prevalent $\pi^0$ topology in this study is from charge current (CC)$\pi^0$ where the scattered muon and the proton from the $\Delta$ decay form the vertex, and there are two disconnected electromagnetic showers from $\pi^0$ decays.   The second is from neutral current (NC)$\pi^0$ where one photon converted within the 0.3 cm wire spacing and thus forms a vertex with the proton, while the second photon is displaced from the vertex. As a result, contributions to the $\pi^0$ selection discussed here will come from NC$\pi^0$ and CC$\pi^0$, as well as CC$\pi^-$ and CC$\pi^+$ events where the $\pi^+/\pi^-$ undergoes charge exchange within the nucleus. 

To fully calculate the kinematics of the $\pi^0$ event we must determine both the energy and the 3D angle of the electromagnetic showers.  This is essential for reconstructing the $\pi^0$ invariant mass—a useful quantity to test the shower reconstruction described in section~\ref{sec:showerreco}. In order to cluster a full 3D shower, the 2D projections on the different wire planes are compared for overlap in time. The overlap fraction is defined as the fraction of shower pixels in the collection plane shower that overlap in time with shower pixels from a 2D shower in another plane, in which the $U$ and $V$ planes are considered separately. If the overlap fraction is $>0.5$ in either or both planes, the pixels that overlap between the collection plane shower and the shower in another plane with the highest overlap fraction are used to calculate a cluster of 3D shower points.  The direction is found by using the calculated center of the 3D point cluster and the event vertex.

This 3D reconstruction leads to what is referred to as the ``$\pi^0$ pre-selection cuts". These are requirements that are necessary in order to reconstruct two 3D showers.
\begin{enumerate}
     \item Vertex passes fiducial volume containment requirement (described in section \ref{sec:initialreco});
     \item Two collection plane showers, each with reconstructed energy greater than 35 MeV;
     \item Both collection plane showers have an overlap fraction with a shower in another plane greater than 0.5;
     \item If a collection plane shower matches with showers in both the $U$ and $V$ planes, the one with the highest overlap fraction is chosen; and
     \item The two collection plane showers cannot match to the same shower in another plane.
\end{enumerate}
After the  "$\pi^0$ pre-selection cuts" have been applied, a substantial number of selected events have showers that are mis-reconstructed. The sample also contains a large number of backgrounds such as cosmic muons. To improve the selection, the following requirements are introduced that reduce mis-reconstruction and remove backgrounds. These are referred to as "box cuts" as they are hard cuts designed to remove background events at the tails of the distributions of various variables.

One of the variables used in these "box cuts" is a $\Delta$ mass test variable. This value is calculated for all those events passing the ``$\pi^0$ pre-selection cuts". The 4-vector of the reconstructed showers are used along with the 4-vector of the proton-like prong. The reconstruction of the proton-like prong is discussed in ref. \cite{3Dreco}. These three objects are assumed to have come from a $\Delta$ decay and are therefore used to reconstruct a $\Delta$ rest mass. The tails of this distribution are comprised of mis-reconstructed $\pi^0$ events and cosmic muon backgrounds which allows for another box cut. The box cuts are then:
 \begin{enumerate}
     \item Reconstructed $\pi^0$ mass is less than 400 MeV;
     \item Reconstructed energy of the leading photon is greater than 80 MeV;
     \item The charge sum of all pixels (both track and shower) within 2 cm of the vertex is greater than 250 Q counts;
     \item Leading shower reconstructed angle w.r.t. beam direction is less than 1.5 radians;
     \item The angle between the two photons is less than 2.5 radians; and
     \item $\Delta$ mass test variable is between 1000 and 1400 MeV.
 \end{enumerate}
 
Here, ``leading photon'' refers to the simulated photon with the highest energy. The reconstructed leading shower is the shower with the highest $Q_{sh}$. An additional requirement on the 1e1p Boosted Decision Tree (BDT) of < 0.7 is further added at this stage to maintain blindness to the LEE for this study  as required by the MicroBooNE blindness procedure \cite{Moon:2020kjr}. This BDT is used to select events with one electron and one proton, so this is a relatively small cut that is a  flat $\approx 3 \%$ effect. Events above this threshold may be mis-reconstructed 1e1p events to which we are currently maintaining blindness.

\begin{figure}[h]
\centering
\begin{subfigure}[b]{0.49\textwidth}
         \centering
         \includegraphics[width=\textwidth]{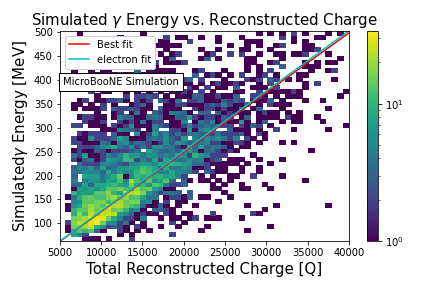}
         \caption{Leading Photon}
     \end{subfigure}
     \hfill
        \begin{subfigure}[b]{0.49\textwidth}
         \centering
         \includegraphics[width=\textwidth]{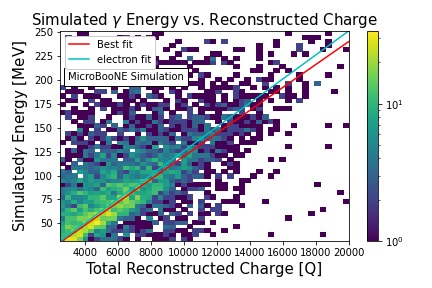}
         \caption{Sub-leading Photon}
     \end{subfigure}

\caption{Simulated photon energy vs $Q_{sh}$ for a sample of generated CC $\pi^0$  events with the $\pi^0$ selection applied. The best fit is shown for this sample as well as the best fit from the electron fit. (a): leading photon, (b): sub-leading photon.
}
\label{ADCtoMeV_gamma}
\centering
\end{figure}

Using the shower energy calculation from the electron sample on both electron and photon showers assumes that the energy of both shower types is reconstructed the same way. To demonstrate that this is valid, the same simulated energy vs $Q_{sh}$ plot and fit is performed on two samples of photons: leading and sub-leading photons from a sample of simulated CC$\pi^0$ events. To ensure this fit is performed only over well-reconstructed events, the $\pi^0$ selection box cuts are applied. The results are shown in figure~\ref{ADCtoMeV_gamma}. The resulting fit equations are: 

\begin{equation}
\label{eq:g1_calib}
    ~\text{Leading Photon: }E~\text{[MeV]} = (1.25 \pm 0.02 \e{-2}) \times Q_{sh}~\text{[$Q$ Counts]}.
\end{equation}

\begin{equation}
\label{eq:g2_calib}
    ~\text{Sub-leading Photon: }E~\text{[MeV]} = (1.20 \pm 0.02\e{-2}) \times Q_{sh}~\text{[$Q$ Counts]}.
\end{equation} 

It is seen from the fit results that the leading photon fit very closely matches the electron fit as expected while the sub-leading photon fit does not match as closely.  This is due to the worse reconstruction in the sub-leading photon. Therefore, the reconstruction uses the charge-to-energy ($Q_{sh}$-to-MeV) conversion value found for the simulation electron sample, $m_{e^-}$.

For the purpose of studying this sample, Monte Carlo simulation has been broken into various categories. "NC $\pi^0$" are neutral current $\pi^0$ events with a well reconstructed vertex, which is a vertex within 5 cm of the true generated vertex. "CC $\pi^0$"  are defined similarly for charged current $\pi^0$. "Offvtx $\pi^0$" are $\pi^0$ events with poorly reconstructed vertices. Here, off-vertex means that the reconstructed vertex is further than 5 cm from a true generated neutrino vertex. "Non $\pi^0$" events are broken into on and off vertex as well. $\nu_e$ events are all events that originated from a $\nu_e$. Cosmic background events also remain after selection.

\begin{figure}[H]
     \centering
     \begin{subfigure}[b]{0.49\textwidth}
         \centering
         \includegraphics[width=\textwidth]{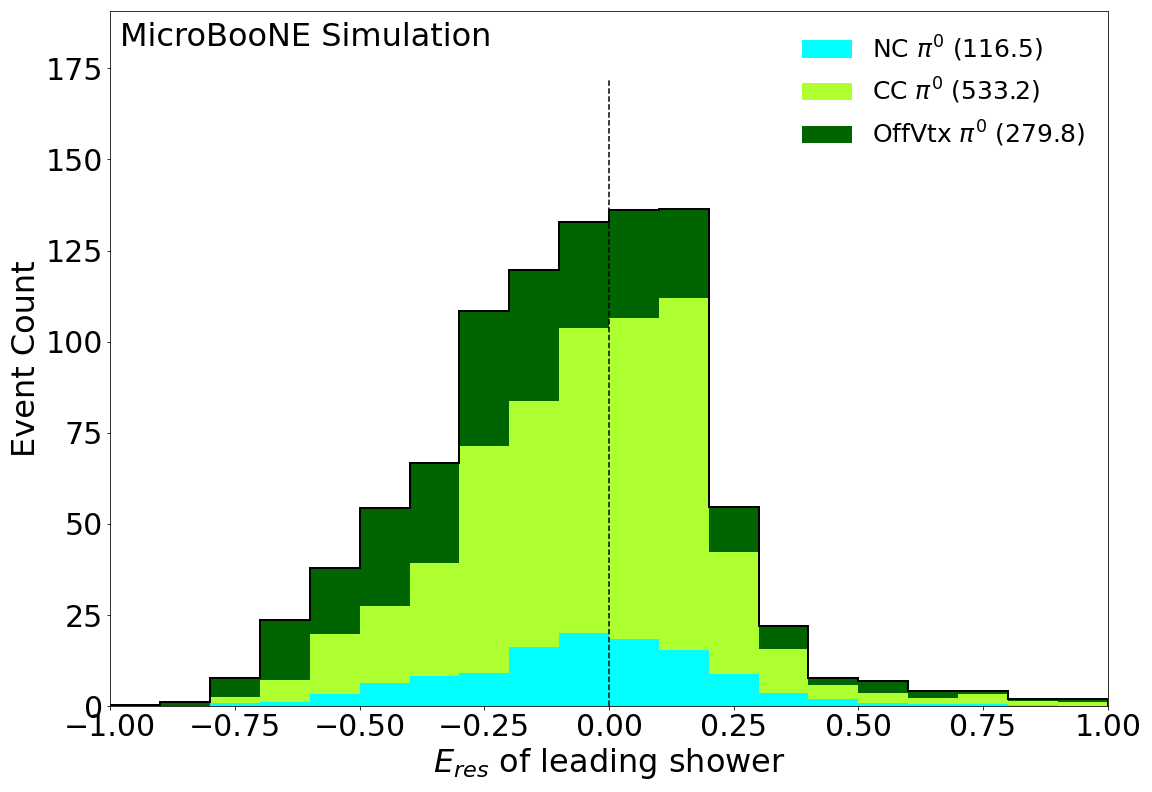}
         \subcaption{Leading Photon}
     \end{subfigure}
     \hfill
     \begin{subfigure}[b]{0.49\textwidth}
         \centering
         \includegraphics[width=\textwidth]{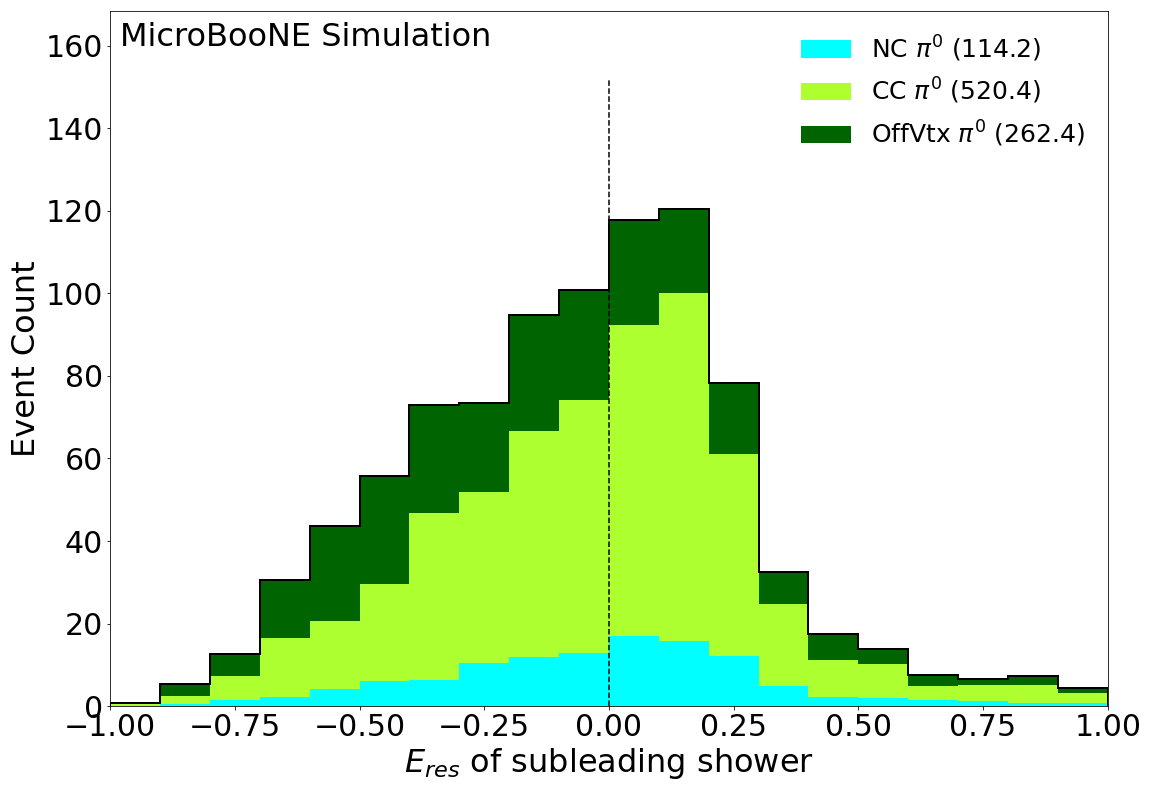}
         \caption{Sub-leading Photon}
     \end{subfigure}
        \caption{The energy resolution for each of the decay photon in the selected $\pi^0$ sample. The leading photon is shown in (a) and the sub-leading photon is shown in (b). The events have been scaled to match the total data POT of $6.67\times10^{20}$. Resolution is defined in eq. \ref{eq:res}. The dashed vertical line is included at $E_{res}=0.0$ for reference.}
        \label{newenergyres}
\end{figure}

Simulated energy resolution of the photons is presented in figure \ref{newenergyres}, where resolution is defined in eq. \ref{eq:res}. This plot is made using the specialized high POT simulation sample of containing only events with a $\pi^0$ in the final state. For the purpose of scaling the different samples and background contributions, the events are POT scaled to match the total data exposure. This distributions in this plot have all of the $\pi^0$ selection cuts applied. The energy resolution is shown separately for leading (highest energy) photon and sub-leading photon. 

Table \ref{tab:energyres} shows the mean and RMS of the distributions shown in figure \ref{newenergyres}. Both the leading and sub-leading photon $E_{res}$ mean is close to zero. The leading photon has a smaller RMS than the sub-leading photon, which is more broad and has a larger tail. The resolution of photons from $\pi^0$ is worse than that seen in the electrons, but they have similar bias as indicated by the mean. There are two main causes for the difference between leading and sub-leading photons. The first is that the sub-leading photon is generally the lower energy of the two. Failing to reconstruct a small number of pixels will have a larger effect on a lower energy shower which has fewer true shower pixels associated with it. The other cause are events where the leading and sub-leading shower are close together. Part of the sub-leading shower is reconstructed as part of the other shower as the leading shower is prioritized. Of these two sub-leading photon reconstruction failure modes, the first is dominant. Some of the other mis-reconstructed photons in both the leading and sub-leading plot are caused by mistakes in SparseSSNet, overlapping cosmic rays that were not removed properly, and unresponsive wires in the collection plane.

\begin{table}[H]
    \centering {
       \caption{ Characterization of the resolution distributions shown in figure \ref{fig:electronres}, figure \ref{newenergyres}, and figure \ref{angleres} .  
    \label{tab:energyres}}
    \begin{tabular}{|l|c|c|c|}
    \hline
    & Mean $E_{sim}$ & Mean $E_{res}$ & RMS of $E_{res}$ \\
    \hline
    Electron & $744.8$ MeV & $-0.07$ & $0.22$\\
    Leading Photon & $230.3$ MeV & $-0.07$ & $0.36$\\
    Sub-leading Photon & $98.8$ MeV & $0.04$ & $0.69$\\
    \hline
     & Mean $\theta_{sim}$ & Mean $\theta_{res}$ & RMS of $\theta_{res}$ \\
    \hline
    Between two photons &  $54.1^{\circ}$ & $0.05$  & $0.50$\\
    \hline
    \end{tabular}}

\end{table}

\begin{figure}[h]
     \centering
     \begin{subfigure}[b]{0.49\textwidth}
         \centering
         \includegraphics[width=\textwidth]{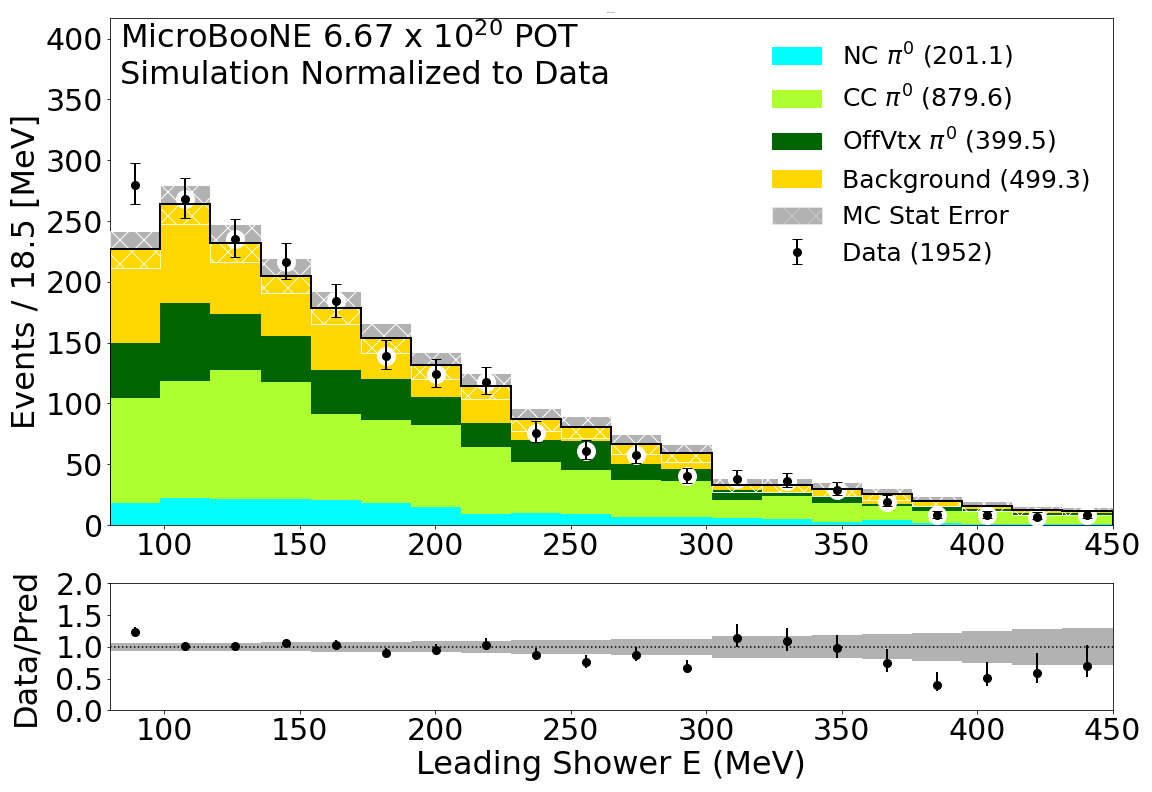}
         \caption{Leading Photon}
     \end{subfigure}
     \hfill
     \begin{subfigure}[b]{0.49\textwidth}
         \centering
         \includegraphics[width=\textwidth]{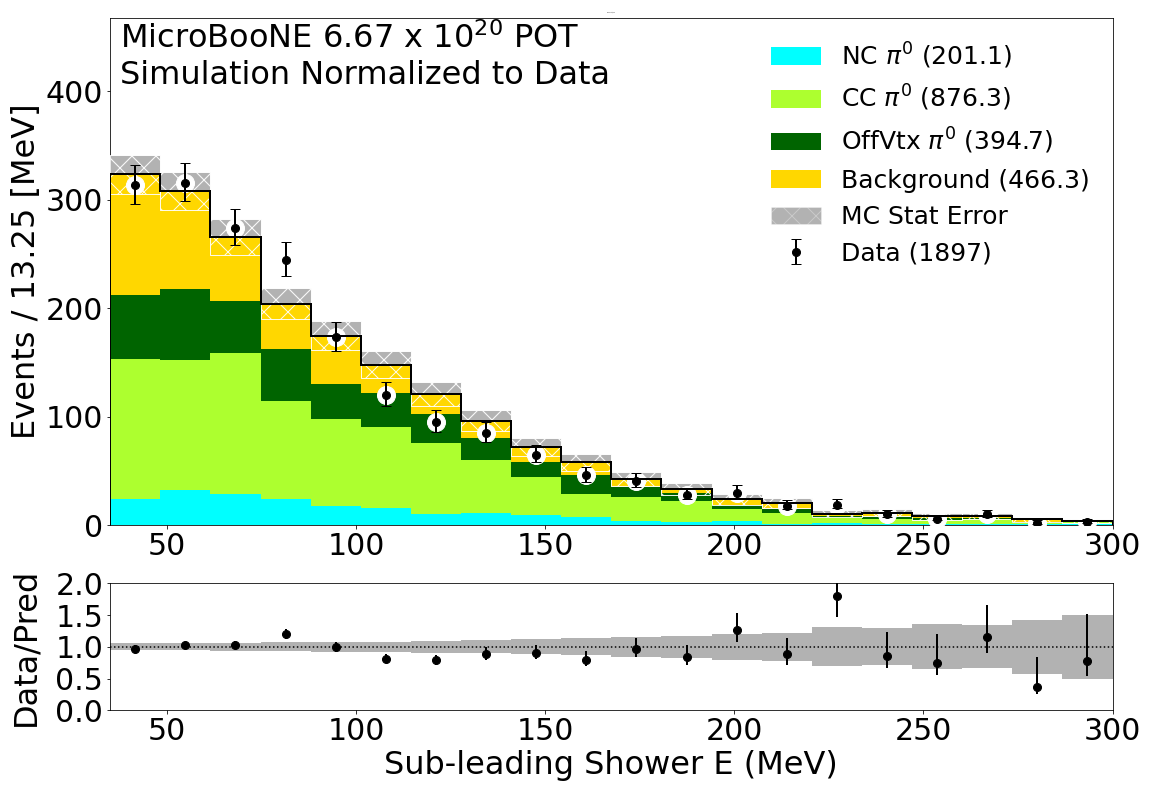}
         \caption{Sub-leading Photon}
     \end{subfigure}
        \caption{The reconstructed photon energies for events passing all selection cuts. The leading photon is shown in (a) and the sub-leading photon is shown in (b). The MC simulation samples have been normalized to the total number of data events. The data events are shown by black points. The number of events in each category is shown in the legend in parenthesises. The $\chi_{CNP}^2/19(dof) = 1.267$ with a p-value of 0.193 for the leading shower and the $\chi_{CNP}^2/19(dof) = 0.973$ with a p-value of 0.491 for the sub-leading shower. }
        \label{fig:pi0showerenergies}
\end{figure}

Figure \ref{fig:pi0showerenergies} shows a data to MC simulation comparison of the reconstructed photon energies in the $\pi^0$ sample. The uncertainty bars on the simulation distribution represent the statistical uncertainty. The total number of simulation events has been scaled to match the total number of data events. This plot uses both the MicroBooNE overlay simulation samples combined with the high POT $\pi^0$ simulation sample and data as described in section \ref{sec:showerreco}. The spectra seen in this figure are at higher reconstructed energies than the Michel $e^-$ in section \ref{sec:michels} and closely mirror the shower energy scale of the MiniBooNE LEE. The $\chi^2$ which is reported in the caption of this plot is a combined Neumann-Pearson (CNP) $\chi^2$~\cite{chi_cnp}. The $\chi^2_{CNP}$ is defined as:

\begin{equation}
\label{eq:cnp}
\chi^2_{CNP} = \sum_i
\begin{cases} 
 \dfrac{(\mu_i- M_i)^2} {\frac{3}{1/M_i + 2/\mu_i}} & M_i \neq 0 \\
\\
 \dfrac{(\mu_i- M_i)^2} {\frac{\mu_i}{2}} & M_i = 0 \\
\end{cases}
\end{equation}
where $\mu_i$ and $M_i$ are the number of predicted and observed events in a given bin.

\begin{figure}[h]
\centering
\includegraphics[width=\textwidth]{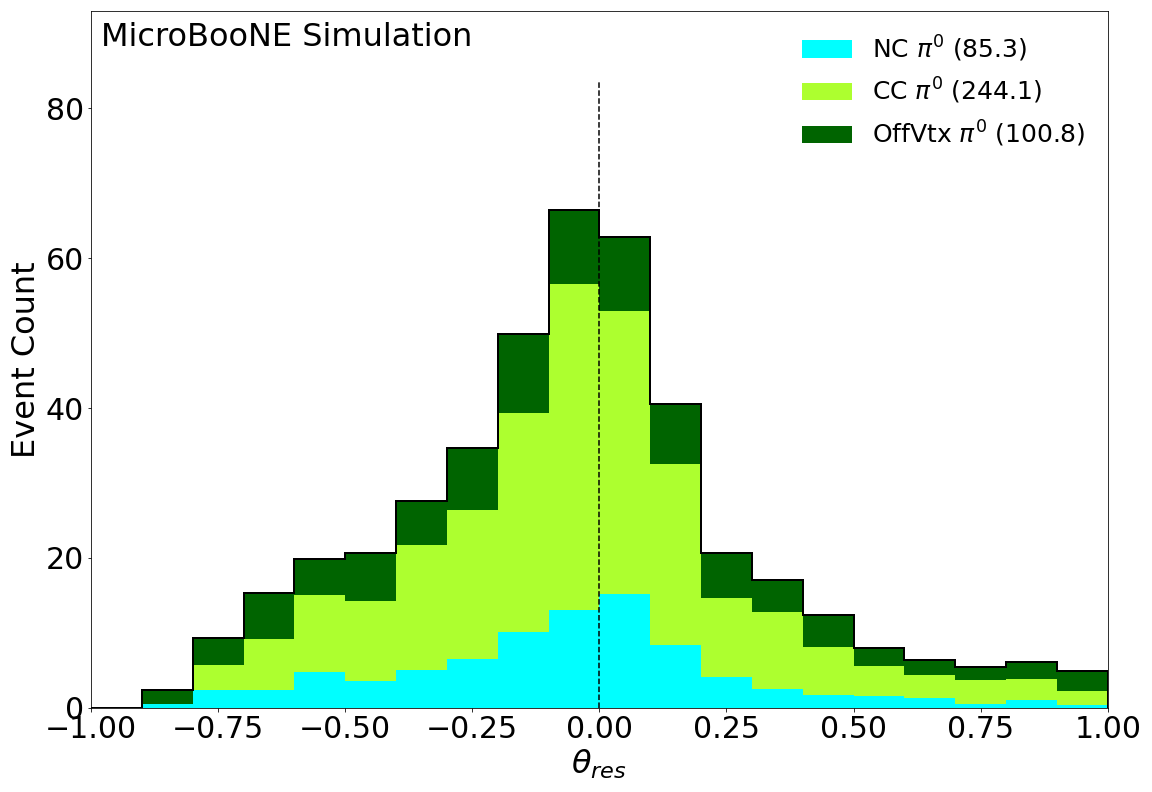}
\caption{The opening angle ($\theta$) resolution of the decay photons in the selected $\pi^0$ sample. The events have been scaled to match total data POT of $6.67\times10^{20}$. Resolution is defined in eq. \ref{eq:angleres}.The dashed vertical line is included at $\theta_{res}=0.0$ for reference.}
\label{angleres}
\end{figure}

Figure \ref{angleres} shows the resolution of  $\theta$.  $\theta$ is the opening angle between the two showers, either simulated or reconstructed, used in the $\pi^0$ mass reconstruction. The opening angle resolution is defined as: 
\begin{equation}
\label{eq:angleres}
    \theta_{res} = \frac{\theta_{reco} - \theta_{sim}}{\theta_{sim}}.
\end{equation}
This figure uses the same simulation sample with the same selection requirements as figure \ref{newenergyres}. The distribution is characterized in Table \ref{tab:energyres}. The mean is close to zero indicating little bias.

\begin{figure}[h]
\centering
\includegraphics[width=\textwidth]{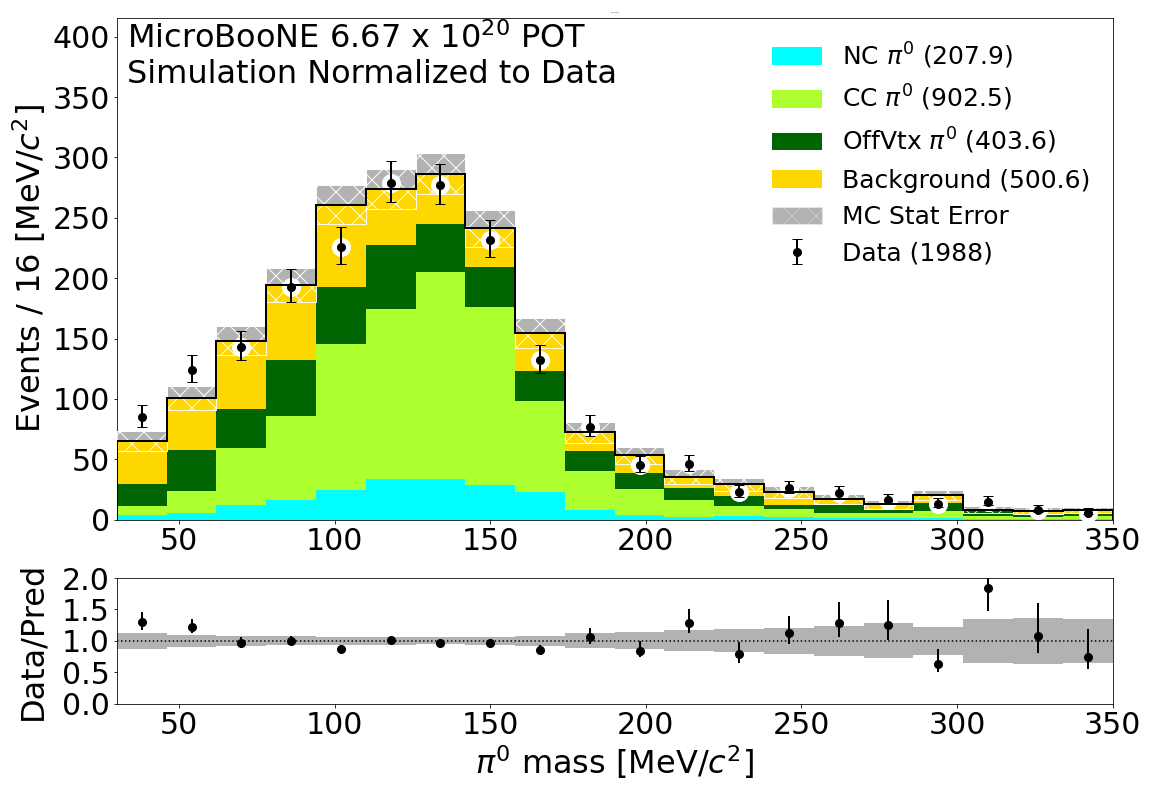}
\caption{The calculated $\pi^0$ mass for events passing all selection cuts. The MC simulation samples have been normalized to total number of data events. The data events are shown by black points. The number of events in each category is shown in the legend in parenthesises. The $\chi_{CNP}^2/19(dof) = 0.976$ with a p-value of 0.486 for the MC prediction. 
}
\label{pi0mass}
\end{figure}

The $\pi^0$ rest mass can now be reconstructed using the following equation:
\begin{equation}
M_\pi^0 = \sqrt{4\sin^2(\frac{\theta}{2})(E_1)(E_2)}
\end{equation}
where $E_1$ is the leading photon energy and $E_2$ is the sub-leading photon energy. The result of this reconstruction is shown in figure \ref{pi0mass}. As in figure \ref{fig:pi0showerenergies}, the total number of simulation events has been scaled to match the total number of data events.  The distributions of both data and simulation peak around 135 MeV, which is the accepted $\pi^0$ rest mass.

\subsection{Validation of Agreement between Simulation, Data, and True Rest Mass}
\label{sec:pi0ADCtoMeVCheck}

The $\pi^0$ sample is next used to verify the agreement in data and simulation of the shower energy scale using the known $\pi^0$ invariant mass $M_{\pi^0}$.  Test points are found representing the best-fit of the $Q_{sh}$-to-MeV conversion factor ($m$) to $M_{\pi^0} = 135$ [$MeV/c^2$].  This is done for a sample of each simulation and data. The goal for each is to find the value of $m$ that yields a $\pi^0$ mass distribution that peaks closest to the true value of 135 [$MeV/c^2$]. This value of $m$ is then compared between data and simulation and to the electron $m_{e^-}$  value found in Section \ref{sec:showerreco}.
 
\begin{figure}[H]
\centering
        \begin{subfigure}[b]{0.49\textwidth}
         \centering
         \includegraphics[width=\textwidth]{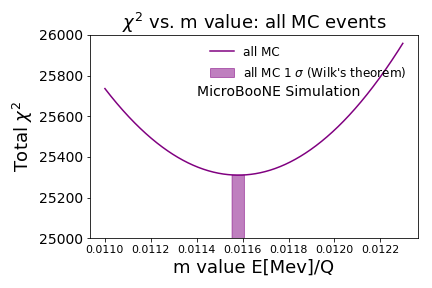}
         \caption{Fit to all selected MC simulation events}
     \end{subfigure}
     \hfill
        \begin{subfigure}[b]{0.49\textwidth}
         \centering
         \includegraphics[width=\textwidth]{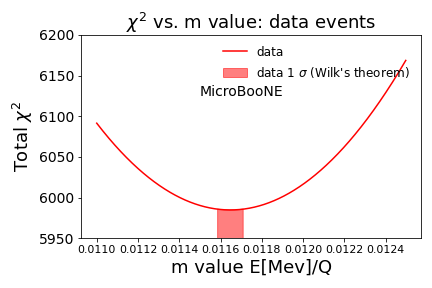}
         \caption{Fit to selected data events}
     \end{subfigure}
     \hfill
      \begin{subfigure}[b]{0.49\textwidth}
         \centering
         \includegraphics[width=\textwidth]{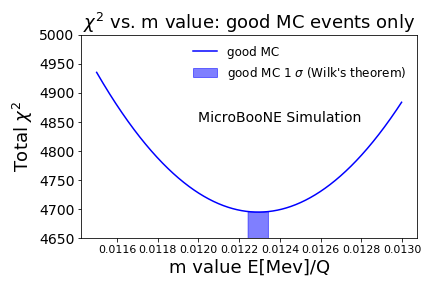}
         \caption{Fit to good selected MC simulation events}
     \end{subfigure}
\caption{Total $\chi^2$ vs. $m$ distributions for  all MC simulation(a), data(b), and good MC simulation(c) that pass the $\pi^0$ selection criteria.
}
\label{pi0_chi2comb}
\end{figure}

 To find the optimal $m$, the following $\chi^2$ formula is minimized:
 
 \begin{equation}
 \label{eq:pi0fit}
\chi^2 = \sum_{i} \bigg( \frac{(135[MeV/c^2] - M^i_{\pi^0})}{dM} \bigg)^2.\
\end{equation}
 where $i$ is each $\pi^0$ event in the given sample, $dM$ is 29.8 [$MeV/c^2$] based on the width of a Gaussian fit to the good simulation $\pi^0$ distribution (the NC $\pi^0$ and CC $\pi^0$ categories in figure \ref{pi0mass}). $M^i_{\pi^0}$ represents the $\pi^0$ mass and is given in this case by:
 \begin{equation}
M^i_{\pi^0} = \sqrt{4\sin^2(\frac{\theta}{2})(m\times(Q_{sh})_1)(m\times(Q_{sh})_2)}
\end{equation}
where $Q_{sh}$ is the reconstructed shower charge and $\theta$ is the reconstructed opening angle between the two showers. The same $\chi^2$ formula is minimized for both MC simulation and data. The 1-$\sigma$ range is calculated on these fit points by looking at the range of $m$ values for each data and simulation which give a $\chi^2$ value satisfying the Wilks' theorem condition $|\chi^2(m) - \min_m[\chi^2(m)]| < 1$ \cite{wilks}. The $\chi^2$ distributions can be found in figure \ref{pi0_chi2comb}. The resulting values can be seen in the first two rows of Table \ref{tab:pi0_mvals}. Excellent agreement is seen between the data and the simulation best fit points indicating that the simulation derived shower conversion factor is valid to use on data.

 Another important consideration is how closely the best fit $m$ values derived here match $m_{e^-}$  in eq. (\ref{eq:orig_calib}). The best fit value of $m$ derived from the $\pi^0$ sample has the potential to be affected by many factors. The largest of these factors is the amount of background events selected. The $\pi^0$ selection presented in section \ref{pi0id} contains many background events, which should not necessarily reconstruct as a $\pi^0$ mass of 135 [$MeV/c^2$]. To account for this background in simulation a second fit is performed only using good simulation events. In this instance good simulation is defined as simulated events that pass all $\pi^0$ cuts, have a simulated final state $\pi^0$, and have a reconstructed vertex within 5 cm of the true simulated neutrino interaction vertex. These are the only types of events in the selection which should result in a reconstructed $\pi^0$ mass of 135 MeV.  A further cut of $\pi^0$ mass $<$ 200 MeV is added to prevent mis-reconstructed events from having a large effect on the $\chi^2$. 
 
 To account for the background events in data, the  optimal $m$ found previously is shifted by the same amount that the MC $m$ is shifted when backgrounds are included ($7.11\times{10^{-4}}$) as seen the last two rows in Table \ref{tab:pi0_mvals}. Data (un-shifted) are the results from minimizing eq. \ref{eq:pi0fit} over all data points. Data (shifted) shows the data points applying the shift found by comparing  the fit over all simulation to the fit over good simulation. This shift retains the excellent data and simulation agreement. The result on data only differs by 1.6\% from the value seen in eq. (\ref{eq:orig_calib}). This indicates that, at the photon energy scale seen in the $\pi^0$ sample, the charge-to-energy conversion factor is valid. Data and simulation agreement of these results and agreement to eq. (\ref{eq:orig_calib}) is discussed further in section \ref{sec:finalcompare} and examined in combination with the results from the Michel $e^-$ sample. 
 
 \begin{table}[H]
\centering
\caption{\label{tab:pi0_mvals}The best value of $m$ (MeV/$Q_{sh}$) for each data and MC simulation sample and the range found using Wilks's theorem. Results are shown before accounting for background (top two rows) and after (bottom two rows).}
\begin{tabular}{ |l | c | c | c| }
   \hline
  Sample & $m$ [MeV/Q] & $m$ range & $\chi^2$/NDF \\ 
  \hline
 All MC & $1.159\e{-2}$ & [$1.156\e{-2}$, $1.160\e{-2}$] & 25310.3/9473 = 2.7 \\
 Data (un-shifted) & $1.165\e{-2}$ & [$1.159\e{-2}$, $1.170\e{-2}$] & 5984.7/1973 = 3.0 \\
 \hdashline

  Good MC & $1.230\e{-2}$ & [$1.225\e{-2}$, $1.234\e{-2}$] & 4694.9/3039 = 1.5\\  
 Data (shifted) & $1.236\e{-2}$ & [$1.230\e{-2}$, $1.241\e{-2}$] & 5984.7/1973 = 3.0
\\
  \hline
\end{tabular}
\end{table}

\section{Michel Electron Sample}
\label{sec:michels}

The second sample used to validate the shower energy reconstruction consists of Michel electrons from decays of $\nu_\mu$-sourced stopped muons in the detector. This is the first time a Michel sample has been reconstructed in a LArTPC which is dominated by Michels from $\nu_\mu$ interactions. As in section ~\ref{sec:pions}, we begin by describing the event selection criteria for this sample. Next, we examine the data/simulation agreement in the Michel shower energy spectrum. Finally, we assess the agreement of the Michel sample with the physical Michel cutoff of $m_\mu/2 = 52.8$~MeV through a fit procedure. The data and MC simulation results of this fit agree, validating the use of the $\nu_e$ simulation-derived $Q_{\rm sh}$-to-MeV on data. Both the data and $\nu_e$ simulation fits also show consistency between the simulation-derived $Q_{\rm sh}$-to-MeV conversion value validating the absolute shower energy scale of the DL-based analysis in the low energy region ($\lesssim 50$~MeV).

\subsection{Identification and Reconstruction}
\label{sec:michelsid}

The Michel electron sample has been chosen in order to validate the reconstructed energy scale of lower-energy electrons, slightly below the energy scale of electrons in the low-energy excess search. Previous work has been performed in MicroBooNE using a larger sample than presented here, as seen in ref. \cite{ubMichel}. The study presented here uses a different selection and is designed to test the reconstruction of showers used in the DL low-energy excess search. However, as shown below, our results are consistent with those shown in \cite{ubMichel} in both the reconstructed Michel energy spectrum and corresponding energy resolution.

Muon-Michel vertices are identified through the following requirements:
\begin{enumerate}
    \item Two prongs at the vertex;
    \item Long prong track-length $>$ 100 cm (candidate muon);
    \item Short prong track-length $<$ 30 cm (candidate Michel);
    \item Long track consists of $< 20\%$ SparseSSNet shower-like pixels (candidate muon);
    \item Short track consists of $>  80\%$ SparseSSNet shower-like pixels (candidate Michel); and
    \item $\phi_\mu <$ 0.5 radians. 
\end{enumerate}
where $\phi_\mu$ is the azimuthal angle of the muon with respect to the horizontal plane, where $\phi_\mu = 0.5$\,rad corresponds to downward-going muons. Lastly, in events with more than one selected vertex, we keep only the first vertex. This strategy was chosen as it does not bias the Michel energy spectrum.

\begin{figure}[H]
    \centering
    \includegraphics[width=.9\textwidth]{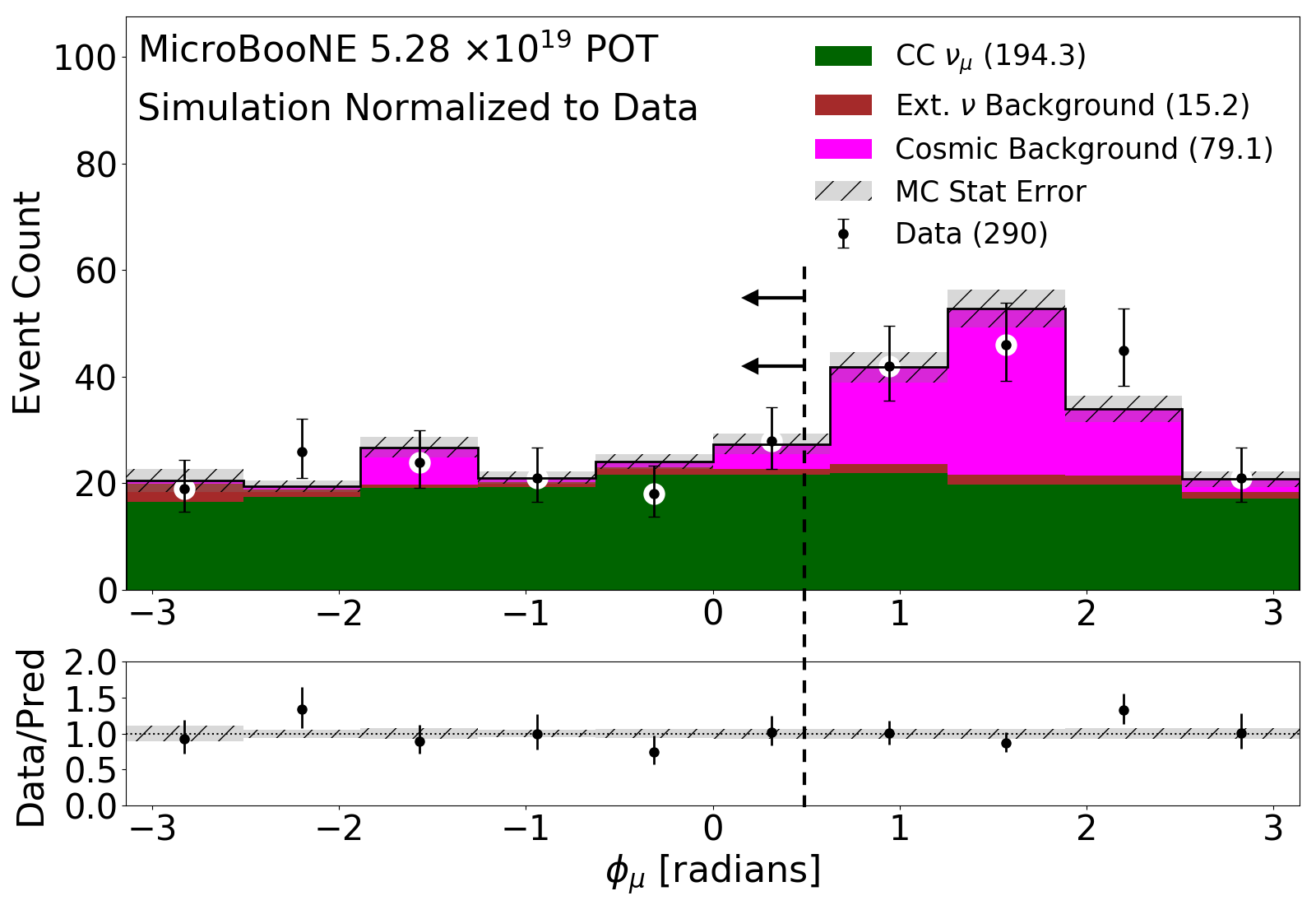}
    \caption{$\phi_\mu$ distribution for selected events in both data and MC simulation, corresponding to $\approx 5.3 \times 10^{19}$ POT. The selection cut requiring $\phi_\mu < 0.5$ radians is indicated by the dotted line. The MC simulation samples have been normalized to total number of data events. The data events are shown by black points. The number of events in each category is shown in the legend in parenthesises. The uncertainty bars here are statistical only.  The $\chi_{CNP}^2/9(dof) = 0.822$ with a p-value of 0.596 for the MC prediction.}
    \label{fig:leptonphicut}
\end{figure}
 
There are two types of Michel electrons that are isolated before the selection. The first are Michels in neutrino events, which can be compared between simulation and data. The second are Michel electrons from stopped cosmic muons. As our simulation samples contain simulated neutrino events overlaid with cosmic data, all of the Michels from stopped cosmic muons come from actual data. Therefore, the only Michels that are truly simulated are those on simulated neutrino events. It is important for these studies to be certain that the Michels in the simulation sample do not come from the stopped cosmic muons. This ensures we are comparing Michels in neutrino events from data to simulated Michels. This is achieved by the final requirement on the muon polar angle $\phi_\mu$, which removes a majority of the predominately downward-going cosmic muons as shown in figure~\ref{fig:leptonphicut}. As discussed in section \ref{sec:initialreco}, the DL vertices search for the intersection of two "prongs". A muon decay into a Michel electron forms this pattern and is therefore often reconstructed at the initial vertex stage of the event reconstruction.

\begin{figure}[H]
    \centering
    \includegraphics[width=.9\textwidth]{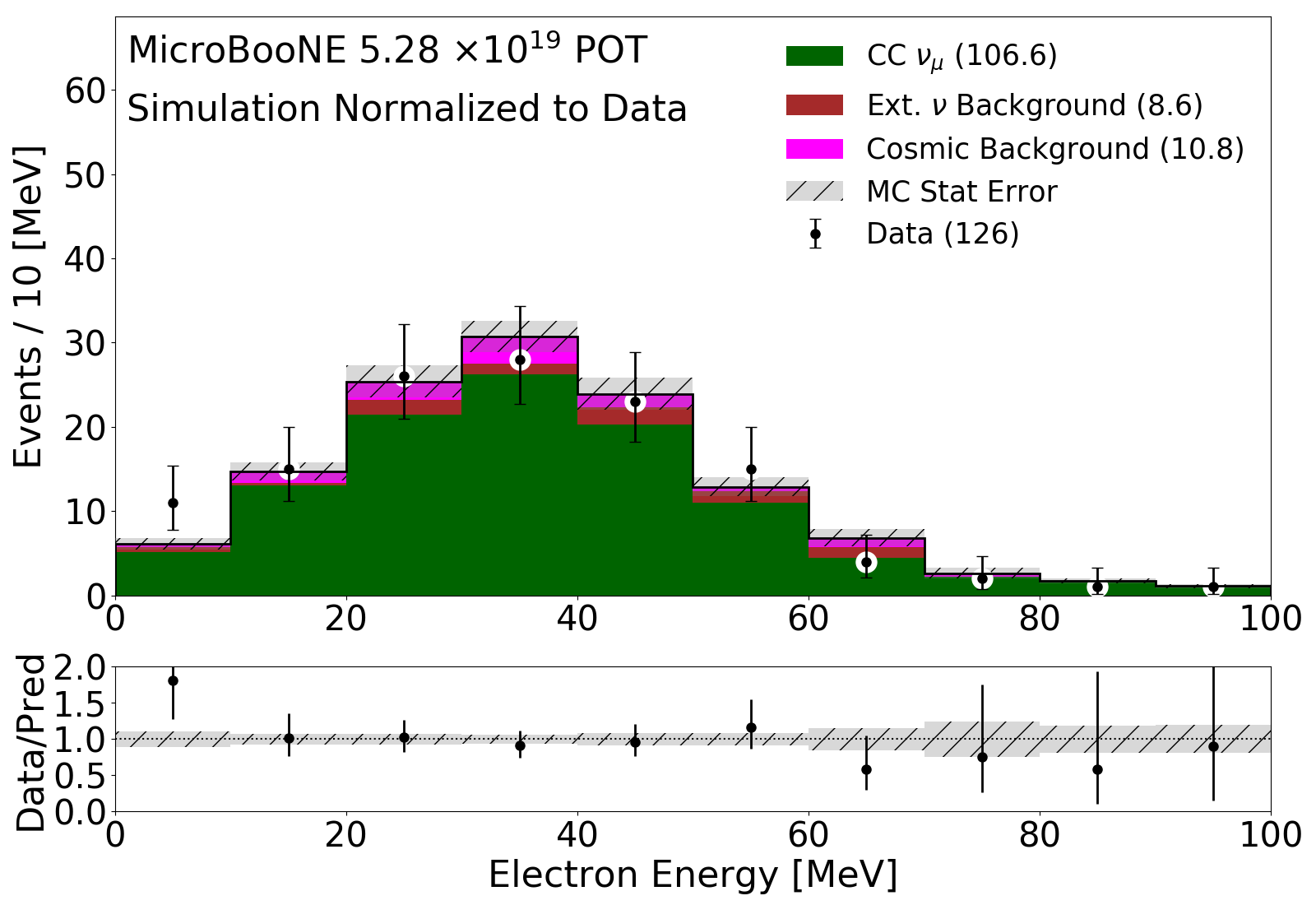}
    \caption{Electron energy distribution for Michels in both data and MC simulation after all selection criteria have been applied, corresponding to $\approx 5.3 \times 10^{19}$ POT. The MC simulation samples have been normalized to total number of data events. The data events are shown by black points. The number of events in each category is shown in the legend in parenthesises. The uncertainty bars here are statistical only.  The $\chi_{CNP}^2/9(dof) = 0.608$ with a p-value of 0.857 for the MC prediction.}
    \label{fig:shower_energy_dataMC}
\end{figure}

For Michel electrons in events passing these cuts, the electromagnetic shower reconstruction algorithm from section \ref{sec:showerreco} is applied to find $Q_{sh}$. This is then converted to a shower energy via eq. \ref{eq:orig_calib}. Figure~\ref{fig:shower_energy_dataMC} shows the shower energy distribution for Michel electrons in both data and simulation. Note that data here come from an open beam data set of corresponding to $\approx5.3\times10^{19}$ POT. 
%This is to ensure blindness to the low-energy excess. 
The simulated events are broken into various categories to indicate which type of event caused the muon. The majority of events come from $\nu_\mu$ interactions within the active detector volume. Some events from cosmic muons and muons from $\nu_\mu$ interactions outside the active volume remain in the sample.

One can see that the high-end tails for the shower energy distribution in both data and simulation fall off around 60 MeV as expected. While not as sharp as the cut-off in ref.~\cite{ubMichel}, the results are consistent. The high energy tail above the true value of $52.8$ MeV likely is due to over estimation of shower energy reconstruction seen in figure \ref{fig:electronres}.  The shower energies of this sample are much lower than those seen in the $\pi^0$ sample. The good data/simulation agreement within statistical uncertainty indicates that the shower algorithm performs well down to low energies. This agreement will be quantified further in the next sections.

\subsection{Validation of Agreement between Simulation, Data, and True Michel Cutoff}
\label{sec:MichelADCtoMeVCheck}

\begin{figure}[H]
    \centering
    \includegraphics[width=\textwidth]{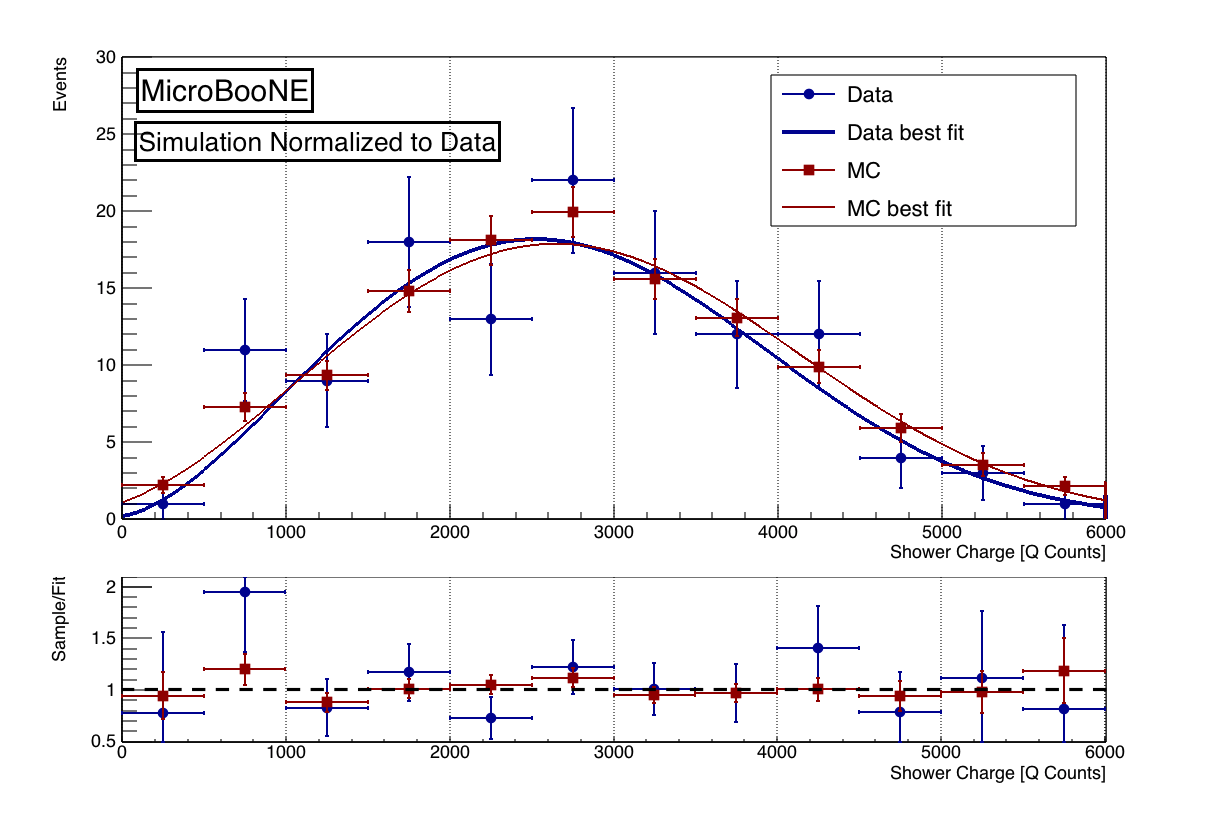} 
    \caption{{\bf Top}: Michel shower charge sum spectrum in data and MC simulation along with the corresponding best fit to eq. (\ref{eq:cutoff_fitter}) (allowing only $Q_\mathrm{cutoff}$ to vary in the fit). This sample corresponds to $\approx 5.3 \times 10^{19}$ POT. {\bf Bottom} Ratio of the data/simulation to the corresponding fit. The MC simulation and fit result here have been normalized to match the data.}
    
    \label{fig:cutoff_bestfit}
\end{figure}

We now perform a validation of the absolute shower energy scale and analyze the data/simulation agreement with the Michel sample analogous to the $\pi^0$ study described in section~\ref{sec:pi0ADCtoMeVCheck}. In the case of the Michel cross-check, we fit the $Q_\mathrm{sh}$ (reconstructed shower charge) spectrum shown in figure~\ref{fig:cutoff_bestfit} to the following five-parameter function:

\begin{equation}
\label{eq:cutoff_fitter}
f(x=\frac{Q_\mathrm{sh}}{Q_\mathrm{cutoff}};\sigma,N) = N \int_0^1 (3y^2 - 2y^3) \frac{1}{\sqrt{2 \pi \sigma^2}} \exp{\frac{-(x-y)^2}{2\sigma^2}}dy 
\end{equation}
where: 
\begin{equation}
\sigma = y\sqrt{r_1^2 + \frac{r_2^2}{y Q_\mathrm{cutoff}} + \bigg(\frac{r_3}{y Q_\mathrm{cutoff}} \bigg)^2}.
\end{equation}

Here, $f(x)$ represents a parameterization of the true Michel spectrum convoluted with a Gaussian representing charge resolution \cite{trueMichel}. $N$ is a floating normalization parameter, and $\{r_1,~r_2,~r_3\}$ represent contributions to the charge resolution corresponding to a constant noise term, a statistical charge-counting term, and a Gaussian noise term, respectively. $Q_\mathrm{cutoff}$ represents the cutoff of the Michel shower energy spectrum, which, after the $Q_\mathrm{sh}$-to-MeV conversion, should correspond to the Michel energy cutoff of $m_\mu/2 \approx 52.8$ MeV. The integration over the variable $y$ represents a scan over the simulated shower charge spectrum. The expression is invariant under $\int_0^1 I(y) dy \to Q_\mathrm{cutoff}^{-1}\int_0^{Q_\mathrm{cutoff}} I(Q_{sh}^* / Q_\mathrm{cutoff}) dQ_\mathrm{sh}^*$, where $I(\dots)$ represents the integrand in eq. (\ref{eq:cutoff_fitter}).

\begin{figure}[H]
     \centering
     \begin{subfigure}[b]{0.49\textwidth}
         \centering
         \includegraphics[width=\textwidth]{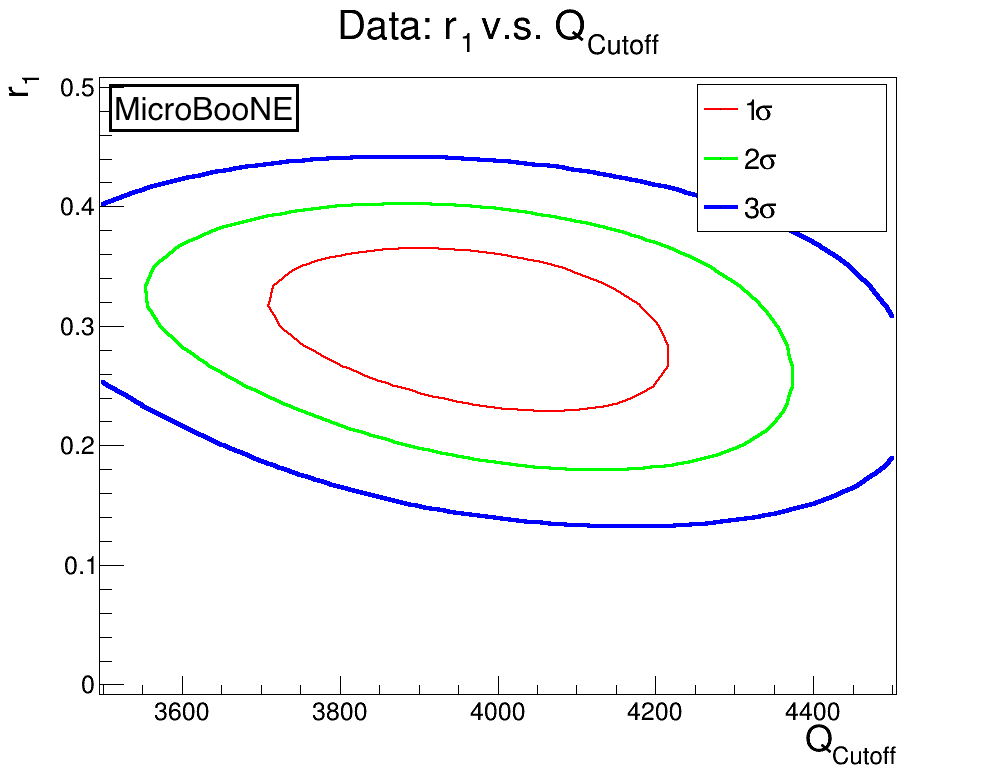}
         \caption{}
         %\caption{Data: $r_1$ v.s. $Q_\mathrm{cutoff}$}
     \end{subfigure}
     \hfill
      \begin{subfigure}[b]{0.49\textwidth}
         \centering
         \includegraphics[width=\textwidth]{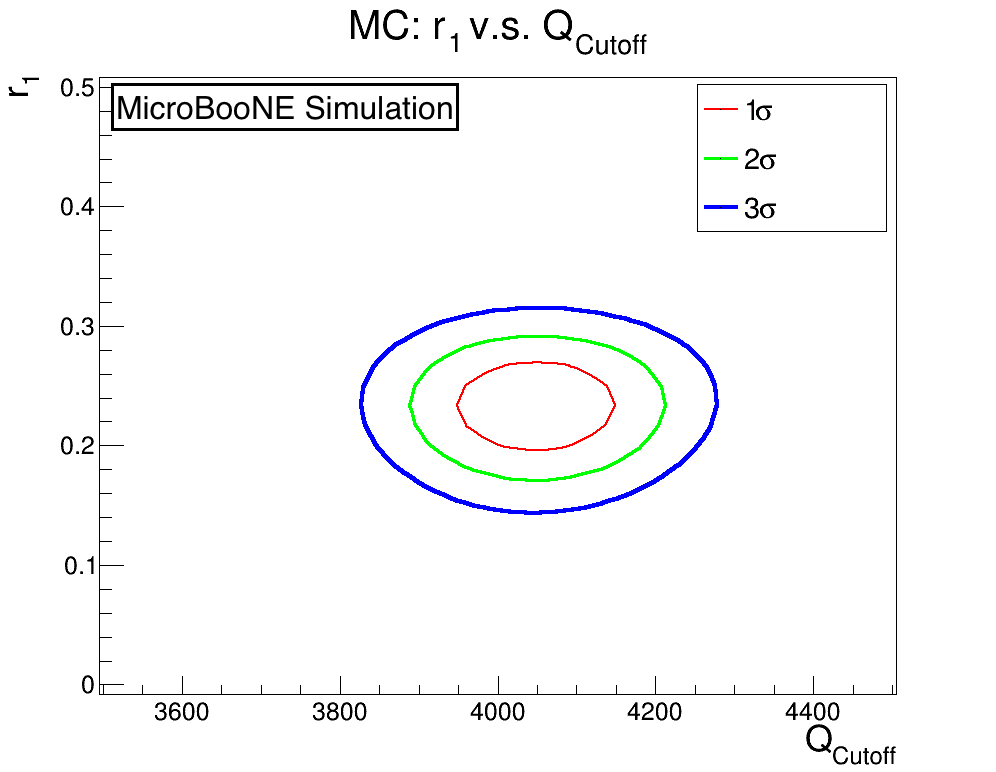}
         \caption{}
         %\caption{MC: $r_1$ v.s. $Q_\mathrm{cutoff}$}
     \end{subfigure}
     \hfill
     \begin{subfigure}[b]{0.49\textwidth}
         \centering
         \includegraphics[width=\textwidth]{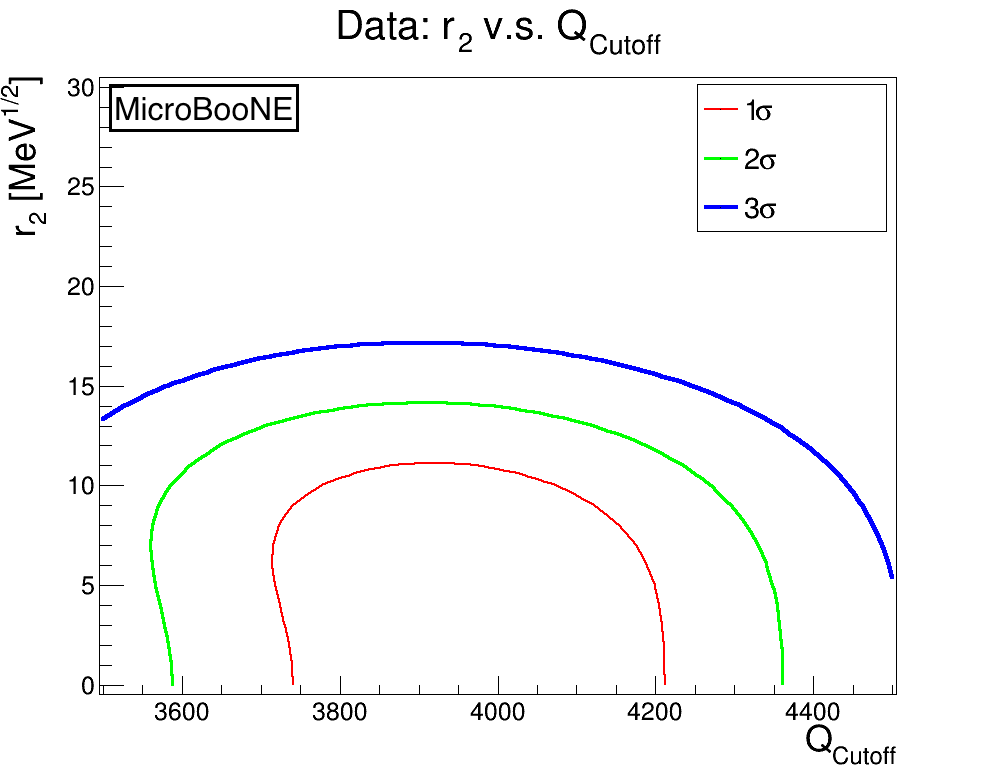}
         \caption{}
         %\caption{Data: $r_2$ v.s. $Q_\mathrm{cutoff}$}
     \end{subfigure}
     \hfill
      \begin{subfigure}[b]{0.49\textwidth}
         \centering
         \includegraphics[width=\textwidth]{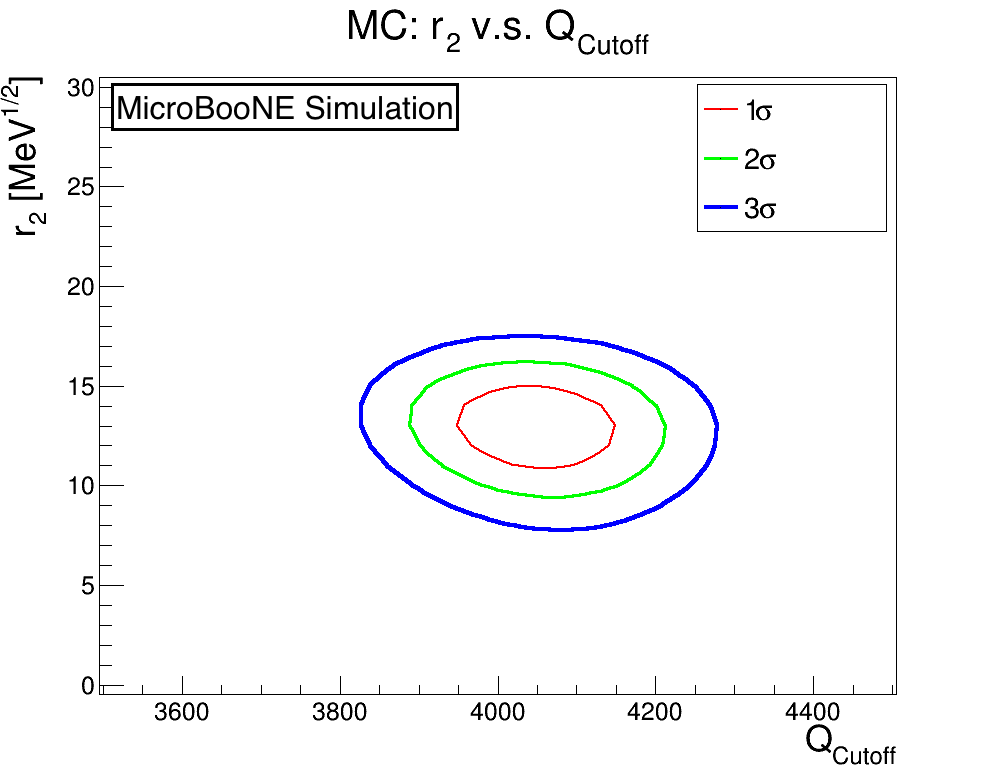}
         \caption{}
        % \caption{MC: $r_2$ v.s. $Q_\mathrm{cutoff}$}
     \end{subfigure}
     \hfill
     \begin{subfigure}[b]{0.49\textwidth}
         \centering
         \includegraphics[width=\textwidth]{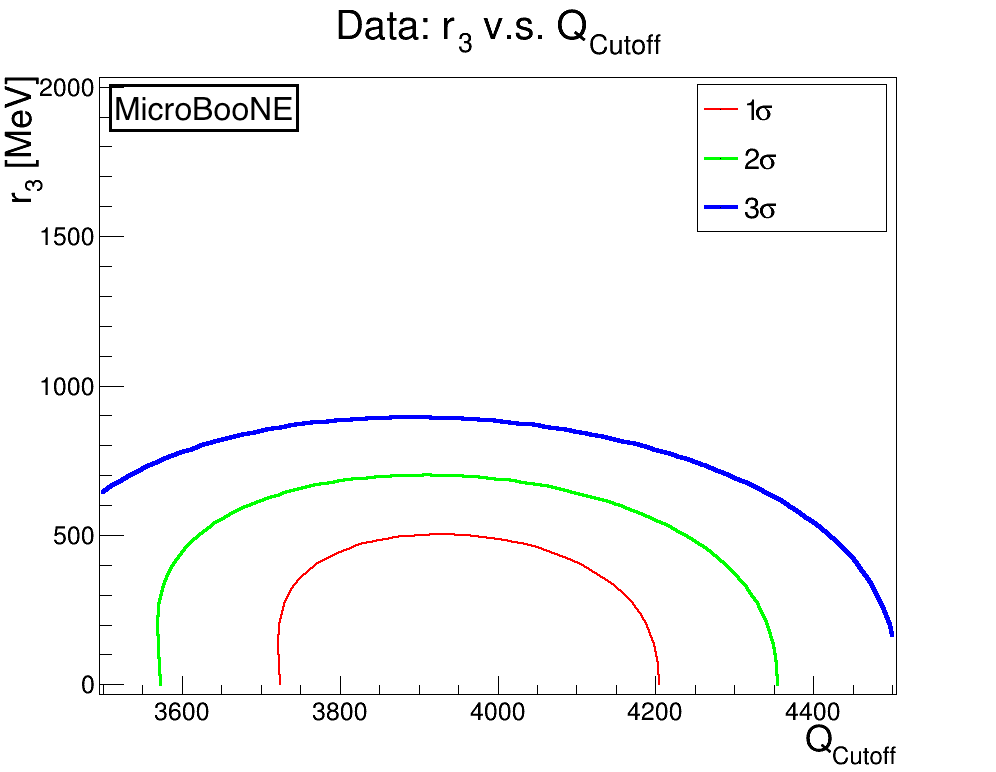}
         \caption{}
         %\caption{Data: $r_3$ v.s. $Q_\mathrm{cutoff}$}
     \end{subfigure}
    \hfill
     \begin{subfigure}[b]{0.49\textwidth}
         \centering
         \includegraphics[width=\textwidth]{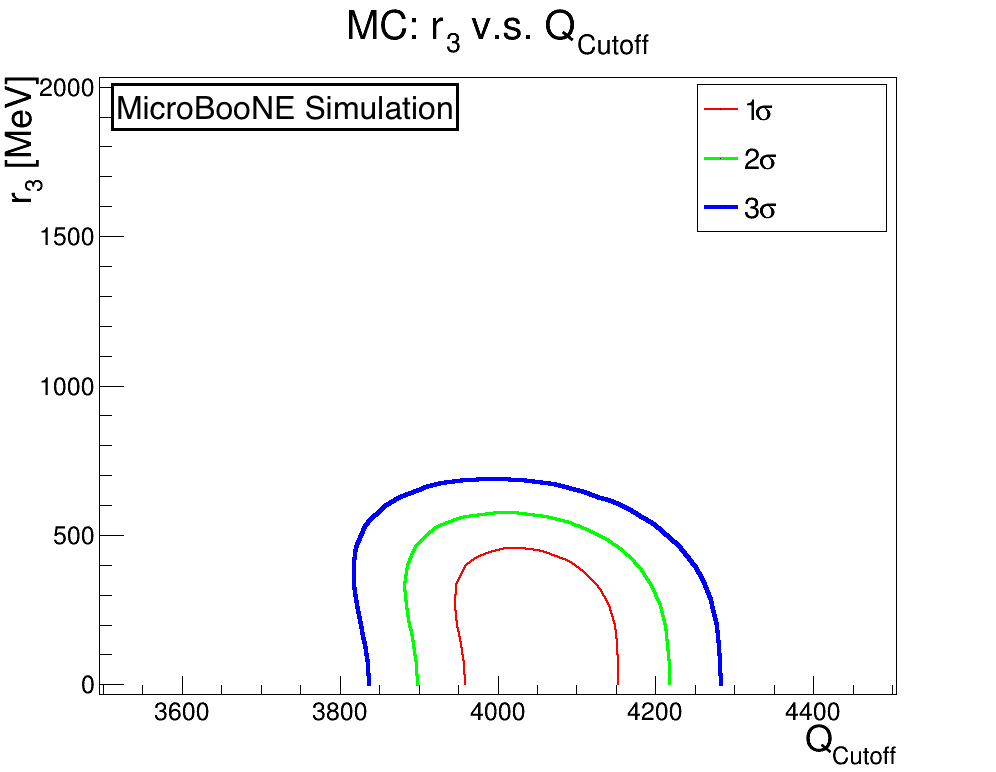}
         \caption{}
         %\caption{MC: $r_3$ v.s. $Q_\mathrm{cutoff}$}
     \end{subfigure}
        \caption{Two-dimensional confidence regions for each resolution parameter in eq. (\ref{eq:cutoff_fitter}) v.s. $Q_\mathrm{cutoff}$. 1$\sigma$, 2$\sigma$, and 3$\sigma$ regions are shown by red, green, and blue curves, respectively. The confidence regions for (a), (c), and (e) come the fit to data while those in (b), (d), and (f) come from the fit to simulation (MC). The units of each parameter in the plots are as follows: $Q_\mathrm{cutoff}$ [Q counts], $r_1$ [dimensionless], $r_2$ [Q counts]$^{1/2}$, $r_3$ [Q counts].}
        \label{fig:2d_michel_data}
\end{figure}

We first fit $f(x)$ to the Michel spectrum in data and simulation by varying all five parameters $Q_\mathrm{cutoff},N,r_1,r_2,r_3$. This is done by minimizing the $\chi^2$:

\begin{equation} \label{eq:cutoff_csq}
\chi^2(Q_\mathrm{cutoff},N,r_1,r_2,r_3) = \sum_{i} \bigg( \frac{(O_i - f(x=\frac{(Q_\mathrm{sh})_i}{Q_\mathrm{cutoff}};N,r_1,r_2,r_3)}{\sigma_\mathrm{i,stat.}} \bigg)^2
\end{equation}
where $O_i$ is the number of observed Michel events in $Q_\mathrm{sh}$ bin $i$ and $\sigma_\mathrm{i,stat.} = \sqrt{O_i}$ is the Poisson error. There are 12 shower charge bins ranging from $0$ – $6000$ counts in this fit. Figure~\ref{fig:2d_michel_data} shows 2D confidence regions for $Q_\mathrm{cutoff}$ versus the different resolution parameters. They are calculated by fixing the remaining three parameters at their best fit values and using Wilks's theorem for two free parameters. As shown in figure~\ref{fig:2d_michel_data}, the fit generally prefers a large contribution from the flat resolution term $r_1$. In data, one can see that the $1\sigma$ $r_1$ contour prefers a fractional resolution of $\approx0.3$, while the $1\sigma$ $r_2$ and $r_3$ contours are both consistent with zero. In simulation, the $1\sigma$ $r_1$ contour prefers a fractional resolution of $\approx0.25$. The $1\sigma$ $r_3$ contour is consistent with zero, but the $1\sigma$ $r_2$ contour prefers a value of $\approx13~$[Q counts]$^{1/2}$. This turns out to be a similar contribution when compared to the flat $r_1$ term. A near-flat energy resolution for Michel showers is consistent with previous MicroBooNE work~\cite{ubMichel}.

\begin{figure}[H]
    \centering
    \begin{subfigure}[b]{0.49\textwidth}
        \includegraphics[width=\textwidth]{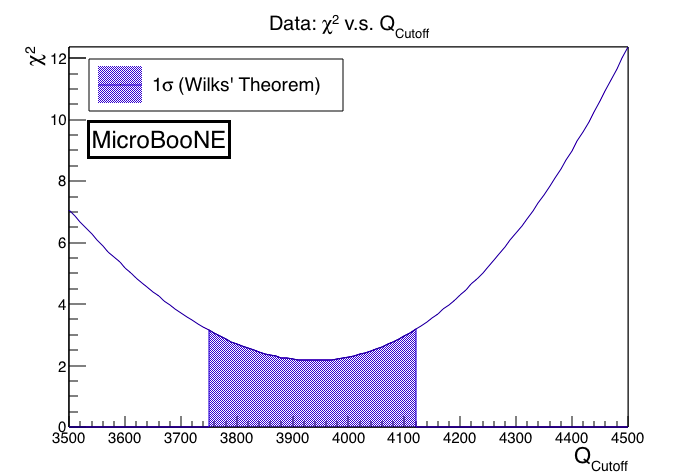}
        \caption{}
    \end{subfigure}
    \hfill
    \begin{subfigure}[b]{0.49\textwidth}
        \includegraphics[width=\textwidth]{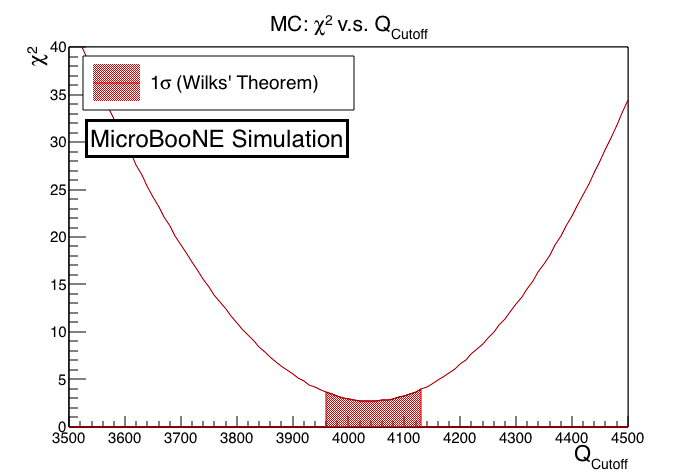}
        \caption{}
    \end{subfigure}
    \hfill
    \caption{$\chi^2$ from (\ref{eq:cutoff_csq}) as a function of $Q_\mathrm{cutoff}$, for data (a) and MC simulation (b). This sample corresponds to $\approx 5.3 \times 10^{19}$ POT. The 1 $\sigma$ allowed regions from Wilks' theorem are shown in shaded regions below each curve.}
    \label{fig:cutoff_chisq}
\end{figure}

Next, the parameters $\{N,r_1,r_2,r_3\}$ are fixed at the minimum $\chi^2$ values (respectively for data/simulation) and a $\chi^2$ scan over $Q_\mathrm{cutoff}$ is performed, this time only including the tail of the $Q_\mathrm{sh}$ spectrum ($Q_\mathrm{sh}$ >3000 Q counts). The purpose of this one-dimensional scan is to obtain the $\chi^2$ minimum and corresponding $1\sigma$ interval on $Q_\mathrm{cutoff}$ using Wilks's theorem for one fit parameter~\cite{wilks}. Figure~\ref{fig:cutoff_bestfit} shows the observed Michel shower charge spectra in data and simulation along with their respective best fits from the 1D scan. Figure~\ref{fig:cutoff_chisq} shows $\chi^2$ as a function of $Q_\mathrm{cutoff}$. The Wilks's theorem 1$\sigma$ interval on $Q_\mathrm{cutoff}$ corresponds to the points for which $\chi^2(Q_\mathrm{cutoff}) - \min\{\chi^2;Q_\mathrm{cutoff}\} \leq 1$. In order to get the charge to energy conversion factor $m$ from $Q_\mathrm{cutoff}$ , $m = \frac{52.8 ~\text{MeV}}{Q_\mathrm{cutoff}}$. The best fit and 1$\sigma$ intervals for $m$ are given in Table~\ref{tab:michel_mvals} along with the $\chi^2$/NDF of the best fit. One can see excellent agreement between data and simulation, demonstrating the consistency of the shower reconstruction. Note that the $1\sigma$ interval on $Q_\mathrm{cutoff}$ is larger in data than in simulation—this is because the statistical error on the data Michel sample is larger than that on the simulated Michel sample. The data/simulation agreement and consistency with eq. (\ref{eq:orig_calib}) demonstrated by this study are discussed further in section~\ref{sec:finalcompare} in combination with the results from the $\pi^0$ sample.

\begin{table}[h]
\centering
\caption{\label{tab:michel_mvals}The best fit values and $1\sigma$ ranges (via Wilks' theorem) for $m$ along the $\chi^2$/NDF of that fit given by eq. (\ref{eq:cutoff_csq}) for both data and  MC simulation. The fit here is the one-dimensional scan over $Q_\mathrm{cutoff}$ transformed into $m$ as described in the text.}
\begin{tabular}{ |l | c | c | c| }
\hline
 Sample  & $m$ [MeV/Q] & $m$ range & $\chi^2$/NDF \\ 
  \hline
 Data &  $1.341\e{-2}$ & [$1.282\e{-2}$, $1.401\e{-2}$] & 2.17/6 = 0.4\\ 
 MC  & $1.308\e{-2}$ & [$1.279\e{-2}$, $1.334\e{-2}$] & 2.73/6  = 0.5\\
 \hline
\end{tabular}
\end{table}

\section{Combined Validation of Reconstructed Shower Energy }
\label{sec:finalcompare}

\begin{figure}[h]
    \centering
        \includegraphics[width=\textwidth]{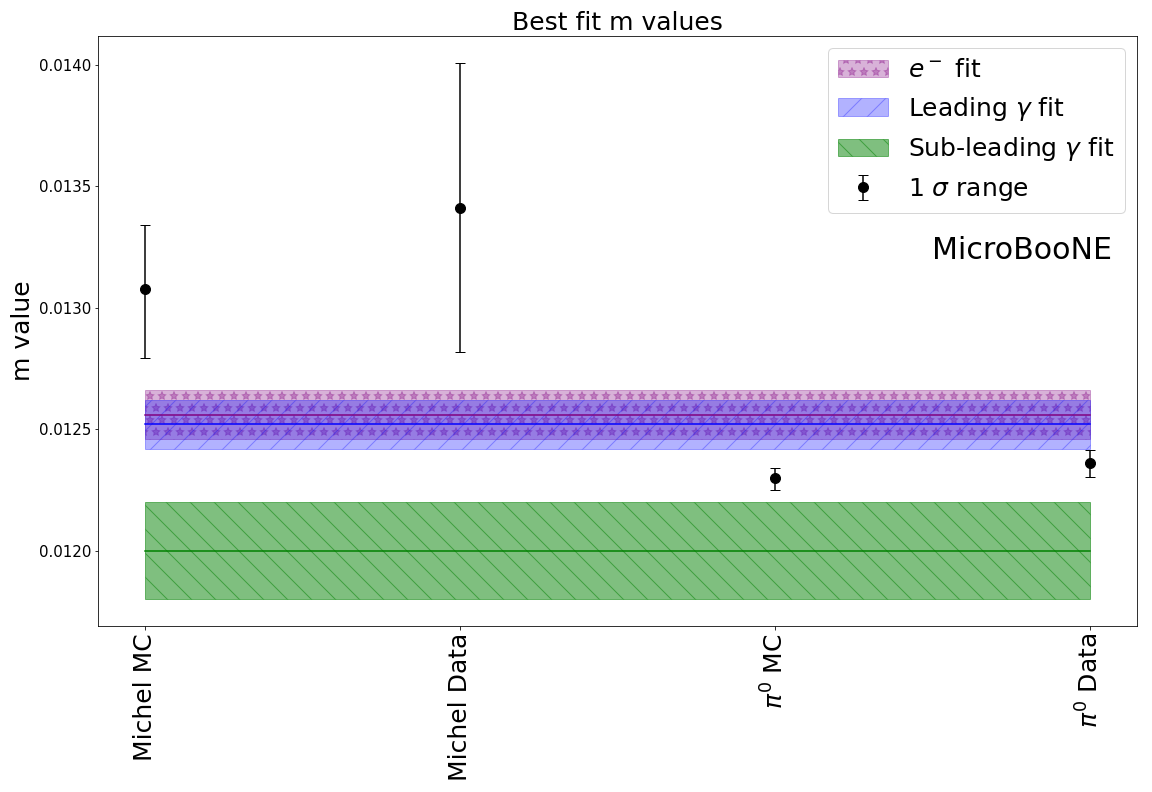}
    \caption{The data and MC simulation points from each the $\pi^0$ sample and the Michel $e^-$ sample are compared with the $Q_\mathrm{sh}$-to-MeV electron calibration line used in the DL analysis from eq. \ref{eq:orig_calib}. The $Q_\mathrm{sh}$-to-MeV photon calibration lines (eq. \ref{eq:g1_calib} and eq. \ref{eq:g2_calib}) are also included for reference. The shaded regions represent the statistical uncertainty of this given calibration line.}
    \label{fig:test_points}
\end{figure}

Both the data/simulation agreement of the shower reconstruction and the absolute scale of the $\nu_e$ simulation-derived $Q_{sh}$-to-MeV conversion are validated using our two samples by utilizing the $\pi^0$ invariant mass of $\approx$135 MeV and the Michel electron spectrum cut-off at $\approx$52.8 MeV. As described in sections \ref{sec:pi0ADCtoMeVCheck} and \ref{sec:MichelADCtoMeVCheck}, we have obtained comparison points separately for data and simulation in each sample. These points are shown with statistical uncertainties in figure~\ref{fig:test_points}. The $Q_{sh}$-to-MeV conversion factor or $m$ values found in section \ref{sec:showerreco} for electrons, leading photons, and sub-leading photons are shown by shaded bands in figure~\ref{fig:test_points}. In principle, one expects agreement between the points and the electron and leading photon calibration line. The 1$\sigma$ ranges in the $m$ value from both the data/simulation Michel cutoff study and the data/simulation $\pi^0$ mass study agree well with the best-fit $m$ values from simulated electrons and leading photons. Agreement with the sub-leading photon line is not necessarily expected because of the reconstruction failure cases discussed previously. 

Table ~\ref{tab:finalagreement} shows the agreement of data and MC simulation for each point, as well as the agreement of each data and simulation point to the electron best fit value from eq. (\ref{eq:orig_calib}). It is seen here that the best fit $m$ values agree between data and simulation for each sample. This validates the use of the same simulation-derived $Q_\mathrm{sh}$-to-MeV conversion value ($m_{e^-}$) for both data and simulation. While the best fit $m$ values for each sample do not exactly match $m_{e^-}$ within statistical uncertainty, there are factors that may affect this value. These include: detector response modeling, sub-leading photon reconstruction in the $\pi^0$ sample, and backgrounds in the Michel $e-$ sample. Therefore, the 2-6\% difference gives an estimate of the scale of the possible data to simulation bias on the shower energy reconstruction. This size effect is acceptable for use in the DL LEE investigation. The $<6.5\%$ difference between each ($m_{MC}$, $m_{Data}$) and $m_{e^-}$ gives the scale of the detector systematic uncertainty in this reconstruction process.

\begin{table}[h]
    \centering
     \caption{Data and MC simulation best fit $m$ values from each sample and comparison to the charge-to-energy conversion factor from eq. (\ref{eq:orig_calib}) ($Q_\mathrm{sh}-to-MeV = m_{e^-} = 1.26 \pm 0.01 \e{-2}$). Uncertainties in ratios are calculated from the 1 $\sigma$ range of each value. The background adjusted values are used for the $\pi^0$ sample.}
     \renewcommand{\arraystretch}{1.5}
    \begin{tabular}{|c|c|c|c|c|c|}
    
        \hline
         Sample & $m_{MC}$ [MeV/Q]& $m_{Data}$ [MeV/Q] & $m_{Data}$/$m_{MC}$ &  $m_{MC}$/$m_{e^-}$ & $m_{Data}$/$m_{e^-}$\\
        \hline
         $\pi^0$ & $1.230^{+0.004}_{-0.006}\e{-2}$ & $1.236^{+0.005}_{-0.006}\e{-2}$ & $1.005 ^{+0.006}_{-0.006}$ & $0.984^{+0.009}_{-0.009}$ & $0.979^{+0.008}_{-0.009}$ \\ 
         \hdashline
         Michel $e^-$ &$1.31^{+0.03}_{-0.02}\e{-2}$ & $1.34^{+0.06}_{-0.06}\e{-2}$ & $1.025^{+0.051}_{-0.049}$ & $1.038^{+0.022}_{-0.024}$ & $1.064^{+0.048}_{-0.048}$ \\

         \hline
         
    \end{tabular}
  
    \label{tab:finalagreement}
\end{table}

In addition, as shown in figure \ref{fig:pi0showerenergies} and figure~\ref{fig:shower_energy_dataMC}, the showers found in the two samples are at different energy ranges. The assumption has been made in this analysis that the $Q_\mathrm{sh}$-to-MeV factor does not change with shower energy. The samples cover the range of values of interest for the DL 1e1p analysis. We conclude that the $Q_{sh}$-to-MeV value given in eq.~\ref{eq:orig_calib} is valid for EM showers in both data and simulation at the energy ranges and precision relevant for the MicroBooNE LEE search.

\section{Conclusions}

This article has reported the updated method of electromagnetic shower reconstruction for the MicroBooNE DL-based LEE analysis. Two samples that allow us to validate our shower reconstruction have been presented: photons produced by the decay $\pi^0$s, and Michel electrons produced when a $\nu_\mu$-sourced stopped muon decays.  The reconstruction and selection for each sample was described. The samples show good data/simulation agreement. The shower energy calculation uses a MC $\nu_e$ simulation-derived $Q_{sh}$-to-MeV conversion factor. The absolute scale of the conversion factor, as well as its application to EM showers in both data and simulations, is validated using the $\pi^0$ invariant mass of $\approx$135 MeV and the Michel electron cut-off at $\approx$53 MeV. Excellent data/simulation agreement is seen in this study. The $Q_{sh}$-to-MeV conversion value used in the DL-based analysis is shown to be consistent with both of these physical quantities. The results we present here form the foundation for the MicroBooNE LEE DL-based analysis of $1e1p$ events that will be released in the future.

\acknowledgments

This document was prepared by the MicroBooNE collaboration using the resources of the Fermi National Accelerator Laboratory (Fermilab), a U.S. Department of Energy, Office of Science, HEP User Facility. Fermilab is
managed by Fermi Research Alliance, LLC (FRA), acting under Contract No. DE-AC02-07CH11359. MicroBooNE is supported by the following: the U.S. Department of Energy, Office of Science, Offices of High Energy Physics and Nuclear Physics; the U.S. National Science Foundation; the Swiss National Science Foundation; the Science and Technology Facilities Council (STFC), part of the United Kingdom Research and Innovation; the Royal Society (United Kingdom); and The European Union’s Horizon 2020 Marie Sklodowska-Curie Actions. Additional support for the laser calibration system and cosmic ray tagger was provided by the Albert Einstein Center for Fundamental Physics, Bern, Switzerland.

\end{document}